\documentclass[letterpaper,english,10pt,aps,letter,superscriptaddress,nofootinbib,pre]{revtex4}
\usepackage[T1]{fontenc}
\usepackage[latin9]{inputenc}
\usepackage{verbatim}
\usepackage{amsmath}
\usepackage{amssymb}
\usepackage{graphicx}
\usepackage{esint}

\makeatletter

\pdfpageheight\paperheight
\pdfpagewidth\paperwidth

\@ifundefined{textcolor}{}
{%
 \definecolor{BLACK}{gray}{0}
 \definecolor{WHITE}{gray}{1}
 \definecolor{RED}{rgb}{1,0,0}
 \definecolor{GREEN}{rgb}{0,1,0}
 \definecolor{BLUE}{rgb}{0,0,1}
 \definecolor{CYAN}{cmyk}{1,0,0,0}
 \definecolor{MAGENTA}{cmyk}{0,1,0,0}
 \definecolor{YELLOW}{cmyk}{0,0,1,0}
}

\usepackage{ae,aecompl}

\makeatother

\usepackage{babel}
\begin{document}
\global\long\def\V#1{\boldsymbol{#1}}
\global\long\def\M#1{\boldsymbol{#1}}
\global\long\def\Set#1{\mathbb{#1}}

\global\long\def\D#1{\Delta#1}
\global\long\def\d#1{\delta#1}

\global\long\def\norm#1{\left\Vert #1\right\Vert }
\global\long\def\abs#1{\left|#1\right|}

\global\long\def\grad{\M{\nabla}}
\global\long\def\avv#1{\langle#1\rangle}
\global\long\def\av#1{\left\langle #1\right\rangle }

\global\long\def\ki{k}
\global\long\def\wi{\omega}

\title{Temporal Integrators for Fluctuating Hydrodynamics}

\author{Steven Delong}

\affiliation{Courant Institute of Mathematical Sciences, New York University,
New York, NY 10012}

\author{Boyce E. Griffith}

\affiliation{Leon H. Charney Division of Cardiology, Department of Medicine, New
York University School of Medicine, New York, NY 10016}

\author{Eric Vanden-Eijnden}

\email{eve2@courant.nyu.edu}

\selectlanguage{english}%

\affiliation{Courant Institute of Mathematical Sciences, New York University,
New York, NY 10012}

\author{Aleksandar Donev}

\email{donev@courant.nyu.edu}

\selectlanguage{english}%

\affiliation{Courant Institute of Mathematical Sciences, New York University,
New York, NY 10012}
\begin{abstract}
Including the effect of thermal fluctuations in traditional computational
fluid dynamics requires developing numerical techniques for solving
the stochastic partial differential equations of fluctuating hydrodynamics.
These Langevin equations possess a special fluctuation-dissipation
structure that needs to be preserved by spatio-temporal discretizations
in order for the computed solution to reproduce the correct long-time
behavior. In particular, numerical solutions should approximate the
Gibbs-Boltzmann equilibrium distribution, and ideally this will hold
even for large time step sizes. We describe finite-volume spatial
discretizations for the fluctuating Burgers and fluctuating incompressible
Navier-Stokes equations that obey a discrete fluctuation-dissipation
balance principle just like the continuum equations. We develop implicit-explicit
predictor-corrector temporal integrators for the resulting stochastic
method-of-lines discretization. These stochastic Runge-Kutta schemes
treat diffusion implicitly and advection explicitly, are weakly second-order
accurate for additive noise for small time steps, and give a good
approximation to the equilibrium distribution even for very strong
fluctuations. Numerical results demonstrate that a midpoint predictor-corrector
scheme is very robust over a broad range of time step sizes.
\end{abstract}
\maketitle

\section{Introduction}

Modeling the effects of thermal fluctuations in spatially-extended
systems requires introducing Langevin stochastic forcing terms in
traditional deterministic models \cite{GardinerBook,OttingerBook}.
Stochastic effects arise in fluid dynamics because of the random thermal
motion of the molecules comprising the fluid at the microscopic level.
Stochastic effects are important in flows at micro and nano scales
typical of new nano- and micro-fluidic and microelectromechanical
devices \cite{Nanofluidics_Review,MultiscaleMicrofluidics_Review},
novel materials such as nanofluids \cite{Nanofluids_Review}, and
biological systems such as lipid membranes, Brownian molecular motors,
and nanopores \cite{Nanopore_Fluctuations}. Thermal fluctuations
can be amplified by non-equilibrium effects and affect the macroscale,
as in fluid mixing \cite{FractalDiffusion_Microgravity,DiffusionRenormalization},
propagation of fronts \cite{PropagatingFronts_Review,FrontPropagation_Review},
combustion of lean flames, capillaries \cite{CapillaryNanowaves,StagerredFluct_Inhomogeneous},
and hydrodynamic instabilities \cite{BreakupNanojets,DropletSpreading,DropFormationFluctuations}.
Because they span the whole range of scales from the microscopic to
the macroscopic \cite{DiffusionRenormalization,FractalDiffusion_Microgravity},
fluctuations need to be consistently included in all levels of description,
including continuum descriptions.

Thermal fluctuations can be included in the classical Navier-Stokes-Fourier
equations of fluid dynamics and related conservation laws through
stochastic forcing terms, as first proposed by Landau and Lifshitz.
The original formulation was for compressible single-component fluids
\cite{Landau:Fluid}. However, the methodology can be extended to
other systems such as fluid mixtures \cite{FluctHydroNonEq_Book},
chemically reactive systems \cite{FluctuatingReactionDiffusion},
magnetic materials \cite{FluctHydro_Magnetic}, and others \cite{LebowitzHydroReview}.
The structure of the equations of fluctuating hydrodynamics can be,
to some extent, justified on the basis of the Mori-Zwanzig formalism
\cite{LLNS_Espanol,DiscreteLLNS_Espanol}. The basic idea is to add
a \emph{stochastic flux} corresponding to each dissipative (irreversible,
diffusive) flux \cite{OttingerBook}, leading to a continuum Langevin
model that ensures the correct equilibrium distribution. Specifically,
statistical mechanics tells us that the stationary (invariant) distribution
at thermodynamic equilibrium is the Einstein distribution for isolated
systems, and the Gibbs-Boltzmann distribution for systems in contact
with a thermal bath.

As model equations, here we consider the fluctuating Burgers equation
in one dimension and the fluctuating Navier-Stokes equation in two
and three dimensions. These stochastic conservation laws have non-dissipative
(skew-adjoint) advective terms and dissipative (self-adjoint) viscous
terms, as well as stochastic forcing terms that are in fluctuation-dissipation
balance with the dissipative terms. Spatial discretizations of the
corresponding continuum (functional) operators should preserve these
(anti-)symmetry properties in order to obey a discrete fluctuation-dissipation
balance. In general, constructing such spatial discretizations (coarse-grained
Langevin models) is non-trivial and may be at odds with other considerations
such as deterministic stability. For example, upwind discretizations
commonly used for advection-diffusion equations add artificial dissipation
to the equations, thus violating fluctuation-dissipation balance.

Based on prior work by us and others, we construct spatial discretizations
for the fluctuating Burgers and fluctuating incompressible Navier-Stokes
equations that obey a discrete fluctuation-dissipation balance principle
just like the continuum equations. The main challenge is in constructing
temporal integrators for the resulting large-scale system of stochastic
differential equations. Ideally, the temporal integrators should have
higher-order short-time accuracy but also lead to long-time dynamics
that is in agreement with fluctuation-dissipation balance. Generalizing
temporal integrators that are favored for Langevin equations in low-dimensional
systems (e.g., molecular dynamics) \cite{Hamiltonian_Leimkuhler}
to the equations of fluctuating hydrodynamics would require implicitly
handling the non-linear advective terms. Solving the resulting large-scale
non-linear system of equations is computationally expensive in three
dimensions for the Navier-Stokes equations. Mixed implicit-explicit
Runge-Kutta schemes are used commonly in the deterministic method
of lines, and can be extended to the stochastic setting with little
effort, at least for the case of additive-noise equations \cite{FD_Nematics_Adhikari}.

We derive the conditions for second-order weak accuracy of implicit-explicit
predictor-corrector schemes for additive noise, and construct several
candidate schemes. These semi-implicit schemes use an implicit midpoint
rule for the diffusive and stochastic terms, which has the remarkable
property that it gives the correct equilibrium distribution independent
of the time step size. The non-linear advective terms are handled
explicitly using an Euler predictor and a trapezoidal or midpoint
corrector. We explain how to incorporate the stochastic forcing term
in the resulting predictor-corrector schemes, and numerically study
their performance on the model fluctuating Burgers equation. The numerical
results suggest that the midpoint corrector is particularly robust.
We then extend the spatio-temporal discretization to handle the incompressibility
constraint present in the fluctuating Navier-Stokes equations in two
and three spatial dimensions. We also include a passively-advected
fluctuating scalar field, as encountered in practical problems. The
schemes that we develop and analyze here have already been employed
in Ref. \cite{LLNS_Staggered} to construct a robust numerical solver
for fluctuating incompressible flows, and to simulate the appearance
of giant concentration fluctuations in diffusively mixing fluids \cite{DiffusionRenormalization,FractalDiffusion_Microgravity}.

We begin by reviewing the fluctuating Navier-Stokes equations, and
summarize a rather general formulation of finite-dimensional Langevin
equations that obey a fluctuation-dissipation principle. In Section
\ref{sec:Burgers} we explain in detail how to spatially discretize
the fluctuating Burgers in one dimension so as to obtain a finite-dimensional
system of Langevin equations with the proper structure. We turn our
attention to temporal integrators for solving the resulting large-scale
structured system of stochastic differential equations in Section
\ref{sec:Weak-Accuracy}. In Section \ref{sec:IMEX} we design several
implicit-explicit stochastic Runge-Kutta schemes that are second-order
weakly accurate for additive noise and maintain fluctuation-dissipation
balance even for large time steps. In Section \ref{sec:FluctNS} we
explain how the spatio-temporal discretization developed for the fluctuating
Burgers equation can be generalized to the fluctuating Navier-Stokes
equations with a passively-advected scalar field. The performance
of the proposed schemes in the nonlinear (large fluctuation) setting
is investigated in Section \ref{sec:Numerical-Results}, and some
concluding remarks are given in Section \ref{sec:Conclusions}. Several
technical calculations are detailed in the Appendix.

\subsection{\label{sub:Nonlinearities}Fluctuating Hydrodynamics}

The prototype stochastic partial differential equation (SPDE) of fluctuating
hydrodynamics is the fluctuating Navier-Stokes equation. This equation
approximates the dynamics of the velocity field $\V v(\V r,t)$ of
a simple Newtonian fluid in the isothermal and incompressible approximation,
$\grad\cdot\V v=0$, 
\begin{equation}
\rho\left(\partial_{t}\V v+\V v\cdot\grad\V v\right)=-\grad\pi+\eta\grad^{2}\V v+\grad\cdot\left[\left(2k_{B}T\eta\right)^{\frac{1}{2}}\M{\mathcal{Z}}\right]+\V f\label{eq:LLNS_incomp_v}
\end{equation}
where $\pi$ is the non-thermodynamic pressure, $\rho$ is the (constant)
density, $\eta=\rho\nu$ is the (constant) shear viscosity and $\nu$
is the kinematic viscosity, and $\V f\left(\V r,t\right)$ is an additional
force density such as gravity \cite{FluctHydroNonEq_Book}. Note that
we prefer to use the standard physics notation instead of the differential
notation more common in the mathematics literature, since the noise
is additive and there is no difference between the different interpretation
of stochastic integrals (e.g., Ito vs. Stratonovich). In the momentum
conservation law (\ref{eq:LLNS_incomp_v}), the stochastic momentum
flux is modeled using a white-noise random Gaussian tensor field $\M{\mathcal{Z}}\left(\V r,t\right)$,
that is, a tensor field whose components are independent (space-time)
white noise processes,
\[
\av{\mathcal{Z}_{ij}(\V r,t)\mathcal{Z}_{kl}(\V r^{\prime},t^{\prime})}=\left(\delta_{ik}\delta_{jl}\right)\delta(t-t^{\prime})\delta(\V r-\V r^{\prime}).
\]
Note that in principle the stochastic momentum flux should have the
symmetrized form $\left(k_{B}T\eta\right)^{1/2}\left(\M{\mathcal{Z}}+\M{\mathcal{Z}}^{T}\right)$
\cite{FluctHydroNonEq_Book}; however, for incompressible flow with
constant viscosity this is not necessary \cite{LLNS_Staggered}.

The fluctuating Navier-Stokes equation, like other augmented Langevin
equations of interest \cite{AugmentedLangevin}, obeys a fluctuation-dissipation
principle, as explained more precisely in Section \ref{sec:GLE}.
Specifically, (\ref{eq:LLNS_incomp_v}) is constructed so that, at
thermodynamic equilibrium, the invariant measure (equilibrium distribution)
for the fluctuating velocities with periodic boundaries is the Gibbs-Boltzmann
distribution with a coarse-grained free energy or Hamiltonian given
by the kinetic energy of the fluid, formally, 
\[
P_{\text{eq}}\left(\V v\right)=Z^{-1}\exp\left[-\frac{\int d\V r\,\rho v^{2}}{2k_{B}T}\right]\V{\delta}\left(\int d\V r\,\rho\V v\right)\delta\left(\grad\cdot\V v\right).
\]
This is ensured by constructing the stochastic forcing term so that
its covariance is proportional to the viscous dissipation operator
$\eta\grad^{2}$. The advective operator $\left(\V v\cdot\grad\right)$,
is, at least formally, Hamiltonian \cite{HamiltonianFluid} in nature,
which means that it preserves the equilibrium distribution dictated
by the competition between the dissipative and the stochastic forcing
terms. We argue that these well-known observations about the structure
of the continuum equations should guide the construction of spatio-temporal
discretizations \cite{Hamiltonian_PDEs}.

We have formally written equation (\ref{eq:LLNS_incomp_v}) as an
infinite dimensional stochastic differential equation. However, the
interpretation of the nonlinear term $\V v\cdot\grad\V v$ requires
giving a precise meaning to products of distributions, which cannot
be defined in general and requires introducing some sort of regularization.
An alternative is to define a discrete hydrodynamic field directly
via some form of averaging of the molecular configuration of the fluid,
and to obtain directly a finite-dimensional system of stochastic \emph{ordinary}
differential equations (SODEs) for the discrete variables through
the Mori-Zwanzig formalism \cite{DiscreteLLNS_Espanol,DiscreteDiffusion_Espanol,Diffusion_MoriZwanzig}.
While such an approach has certain advantages from a coarse-graining
perspective, the notion of a continuum equation and the applicability
of traditional methods for computational fluid dynamics is lost or
at least obscured.

Here we adopt a middle ground between the ``continuum'' and the
``discrete'' approach to fluctuating hydrodynamics. Specifically,
we first spatially discretize the SPDE to obtain a system of SODEs,
in the spirit of the ``method of lines''. Our focus here is on the
temporal integrators for the resulting system of SODEs. We do \emph{not}
consider the convergence of the numerical method as the spatial discretization
is refined, as one would in deterministic fluid dynamics. Rather,
we fix the spatial discretization and assume that the hydrodynamic
cells are sufficiently large, specifically, that they contain, on
average, sufficiently many fluid molecules $N_{p}\gg1$. This ensures
that the equilibrium fluctuations will be on the order of $O(N_{p}^{-1/2})$
relative to macroscopic fields. We also assume that the transport
coefficients have been \emph{renormalized} to account for the finite
number of fluid particles (molecules) used to define the hydrodynamic
fields \cite{DiffusionRenormalization_PRL,DiffusionRenormalization}.
There is strong numerical evidence that under these conditions spatio-temporal
discretizations can correctly capture the leading-order (measurable)
effects of fluctuations at large scales, such as fluctuation-driven
transport in non-equilibrium systems \cite{DiffusionRenormalization},
large-scale inhomogeneities arising during free fluid mixing \cite{LLNS_Staggered},
and diffusive effects on the very long-time dynamics such as drifts
in propagating fronts \cite{DiffusingFront_Saarloos} and shocks \cite{Bell:07}.

\subsection{\label{sec:GLE}Fluctuation-Dissipation Balance in Generic Langevin
Equations}

The fluctuating hydrodynamic formalism finds its foundation in the
theory of coarse-graining \cite{CoarseGraining_Pep}. One of the central
objects in the theory is the \emph{coarse-grained Hamiltonian} or
a \emph{coarse-grained free energy}, which determines the Gibbs-Boltzmann
equilibrium probability density for the coarse variables $\V x\in\Set R^{N}$,
\begin{equation}
P_{\text{eq}}\left(\V x\right)=Z^{-1}\exp\left[-\frac{H\left(\V x\right)}{k_{B}T}\right],\label{eq:GibbsBoltzmann}
\end{equation}
where $Z$ is a normalization factor and the scaled temperature $k_{B}T$
sets the unit of energy. We will set $k_{B}T=1$ for simplicity through
the remainder of this paper. Note that in many cases the actual Hamiltonian
(total energy) may be expressible in terms of the coarse-grained variables
and is strictly conserved \cite{OttingerBook}, but this is not the
case for isothermal systems.

As discussed at length by Grabert \cite{GrabertBook}, a Markovian
approximation within the Mori-Zwanzig formalism can be used to obtain
a \emph{generic Langevin equation} of motion \cite{AugmentedLangevin}
for an arbitrary choice of the microscopic ensemble,
\begin{equation}
\partial_{t}\V x=-\M N\left(\V x\right)\frac{\partial H}{\partial\V x}+\left(2k_{B}T\right)^{1/2}\M B\left(\V x\right)\,\M{\mathcal{W}}(t)+\left(k_{B}T\right)\frac{\partial}{\partial\V x}\cdot\M N^{\star}\left(\V x\right),\label{eq:x_t_general}
\end{equation}
where $\M{\mathcal{W}}\in\Set R^{N_{w}}$ denotes white noise, the
formal temporal derivative of a collection of independent Brownian
motions, and an Ito interpretation is assumed. We will typically suppress
the explicit dependence on $\V x$ and write the \emph{mobility operator}
as $\M N\equiv\M N\left(\V x\right)$. In the generic Langevin equation
(\ref{eq:x_t_general}), the skew-adjoint operator 
\[
\M S=-\M S^{\star}=\frac{1}{2}\left(\M N^{\star}-\M N\right)
\]
generates the ``conservative'' part of the dynamics, and the self-adjoint
positive semi-definite operator 
\[
\M M=\M M^{\star}=\frac{1}{2}\left(\M N+\M N^{\star}\right)\succeq\M 0
\]
generates the ``dissipative'' part of the dynamics, since
\[
\frac{dH}{dt}=-\left(\frac{\partial H}{\partial\V x}\right)^{T}\M N\frac{\partial H}{\partial\V x}=-\mbox{Re}\left[\left(\frac{\partial H}{\partial\V x}\right)^{T}\cdot\M M\cdot\frac{\partial H}{\partial\V x}\right]\leq0.
\]
The operator $\M B\left(\V x\right)$ is constrained, but not uniquely-determined,
by the \emph{fluctuation-dissipation balance} condition
\[
\M B\M B^{\star}=\M M.
\]
The last term in (\ref{eq:x_t_general}) is an additional ``spurious''
or ``thermal'' drift term whose form depends on the particular interpretation
of the stochastic equations \cite{KineticStochasticIntegral_Ottinger},
which we take to be in the sense of Ito. In the case of the fluctuating
Burgers and Navier-Stokes equations this term will vanish.

The dynamics (\ref{eq:x_t_general}) is ergodic and time reversible
(under an appropriate parity transformation for the variables) with
respect to the distribution (\ref{eq:GibbsBoltzmann}). It is not
hard to show using the corresponding Fokker-Planck Equation that (\ref{eq:GibbsBoltzmann})
is an equilibrium distribution for (\ref{eq:x_t_general}), as desired,
at least if one makes the nonrestrictive assumption that $\M M\succ\M 0$.
More precisely, the invariant measure for the coarse-grained variables
is $d\mu_{\text{eq}}=\mu_{\text{eq}}\left(d\V x\right)=P_{\text{eq}}\left(\V x\right)d\V x$.
We will assume here that the dynamics is ergodic with respect to a
unique equilibrium distribution, which may not be the case if $H\left(\V x\right)$
is not defined everywhere. It is important to point out that when
Langevin equations of the form (\ref{eq:x_t_general}) are applied
in a non-equilibrium setting, the dynamics may not be ergodic with
respect to any distribution, even at steady state.

\section{\label{sec:Burgers}Fluctuating Burgers Equation}

The analysis and numerical solution of the incompressible Navier-Stokes
equation is complicated by the presence of the incompressibility constraint.
We begin our discussion by constructing a spatial discretization of
the simpler unconstrained \emph{fluctuating Burgers equation} for
the random field $u\left(x,t\right)$,
\begin{equation}
\partial_{t}u+cu\,\partial_{x}u=\nu\,\partial_{xx}^{2}u+\left(2\nu\right)^{\frac{1}{2}}\,\partial_{x}\mathcal{Z},\label{eq:stochastic_Burgers}
\end{equation}
where $\nu$ is a diffusion coefficient and $c$ sets the scale for
the advection speed. This equation mimics some of the properties of
the fluctuating Navier-Stokes equation (\ref{eq:LLNS_incomp_v}),
in particular, it obeys a fluctuation-dissipation balance principle
with respect to the Gibbs-Boltzmann distribution with a Hamiltonian
$H=\int dx\, u^{2}/2$. The fluctuating Burgers equation can also
be written in conservative form
\[
\partial_{t}u=-\partial_{x}\left[c\frac{u^{2}}{2}-\nu\partial_{x}u-\left(2\nu\right)^{\frac{1}{2}}\mathcal{Z}\right],
\]
showing that the total momentum $\int u\, dx$ is conserved with periodic
boundary conditions. Note that here the stochastic forcing term is
linear and involves the spatial derivative of white noise \cite{AMAR_Burgers},
rather than white noise itself as in the stochastic Burgers equation
studied, for example, in Ref. \cite{BurgersApproximations_Hairer}.
Equations of this type arise as coarse-grained models of the behavior
of one dimensional lattice gases, such as the asymmetric excluded
random walk model \cite{AMAR_Burgers}.

In this section we show how the fluctuating Burgers equation can be
spatially discretized in a manner that leads to a generic Langevin
equation of the form (\ref{eq:x_t_general}). This construction will
be extended to the Navier-Stokes equations with a passively-advected
scalar in Section \ref{sec:FluctNS}. Our approach to the spatial
discretization follows standard practice in deterministic fluid dynamics.
Specifically, we construct the spatially-discrete system by combining
locally-accurate spatial discretizations of the differential operators
(e.g., gradient, divergence and Laplacian) that appear in the the
SPDE. However, in addition to focusing on accuracy and stability when
choosing the spatial discretization, we pay particular attention to
preserving \emph{fluctuation-dissipation balance}. This means that
we want to obtain a system of SODEs whose structure is given in (\ref{eq:x_t_general})
and whose invariant distribution (equilibrium distribution) is a natural
discretization of the Gibbs distribution dictated by equilibrium statistical
mechanics.

\subsection{Continuum Fluctuating Burgers Equation}

One can, at least formally, consider a generic Langevin equation for
an infinite dimensional field \cite{OttingerBook}. The fluctuating
Burgers equation (\ref{eq:stochastic_Burgers}) is a prototype of
such an equation. In this formalism the coarse-grained Hamiltonian
is a functional of the field and the partial derivatives should be
interpreted as functional derivatives, and contractions by a field
imply integrations over the spatial domain. For equation (\ref{eq:stochastic_Burgers}),
the (formal) free energy functional is
\begin{equation}
H\left[u\left(x,t\right)\right]=\int\frac{u^{2}}{2}dx,\label{eq:H_Burgers_continuum}
\end{equation}
so that
\[
\frac{\partial H}{\partial u}\equiv\frac{\d{H\left[u\left(x,t\right)\right]}}{\d u}=u.
\]

The dissipative and fluctuating dynamics in (\ref{eq:stochastic_Burgers})
are generated by the constant operators,
\[
\M M=-\nu\partial_{xx}^{2}\mbox{ and }\M B=\nu^{\frac{1}{2}}\partial_{x},
\]
which in higher dimensions become multiples of the Laplacian and divergence
operators, respectively. The conservative dynamics for the Burgers
equation is Hamiltonian and generated by the skew-adjoint linear operator
$\M S\left(u\right)$ defined through its action on a field $w\left(x,t\right)$
\cite{HamiltonianFluid},
\begin{equation}
\M S\left(u\right)\, w=-\frac{c}{3}\left[u\partial_{x}w+\partial_{x}\left(uw\right)\right].\label{eq:L_Burgers}
\end{equation}
The $\V v\cdot\grad\V v$ term in the higher-dimensional fluctuating
Navier-Stokes equation (\ref{eq:LLNS_incomp_v}) can similarly be
written in terms of a skew-adjoint operator, although there are some
complications in handling the divergence-free constraint \cite{GENERIC_Mixtures}.

A detailed description of the meaning and importance of the Hamiltonian
nature of the nonlinear deterministic dynamics and the Poisson bracket
associated with $\M S$ is given in Refs. \cite{HamiltonianFluid,OttingerBook}.
For our purposes, the most important property of Hamiltonian dynamics
is that it is incompressible in phase space,
\begin{equation}
\frac{\partial}{\partial u}\cdot\M S\left(u\right)=\frac{\partial}{\partial u}\cdot\M N^{\star}\left(u\right)=\V 0.\label{eq:div_free_L}
\end{equation}
This implies that the dynamics of the inviscid Burgers equation preserves
not just functions (such as the Hamiltonian itself) but also phase-space
measures (such as the Gibbs distribution), and thus any probability
density that is a function of $H$ only is a candidate equilibrium
distribution. The inviscid Burgers equation may also be written in
Hamiltonian form using the Hamiltonian $H=\int\left(u^{3}/6\right)dx$
with $\M S=-\partial_{x}$ \cite{HamiltonianFluid,Hamiltonian_BurgersHopf}.
However, in fluctuating hydrodynamics the choice of the coarse-grained
Hamiltonian is dictated by statistical mechanics and the equilibrium
Gibbs distribution is maintained via the fluctuation-dissipation balance
between the viscous and stochastic terms.

\subsection{\label{sub:BurgersFD}Discrete Fluctuating Burgers Equation}

The preceding discussion of how the fluctuating Burgers equation can
be written in the form of a generic Langevin equation (\ref{eq:x_t_general})
is formal and merely informs our choice of spatial discretization.
The discretized $\V u=\left\{ u_{1},\dots,u_{N}\right\} $ can be
thought of as a finite-volume representation of the field $u(x,t)$
on a regular grid with spacing $\D x$, specifically, $u_{j}$ can
be thought of as representing the average value of $u(x,t)$ over
the interval (cell) $\left[j\D x,\,(j+1)\D x\right]$. As we already
discussed, this is merely a formal association and the actual physical
object is the discrete (coarse-grained) $\V u\left(t\right)$ and
not the hypothetical $u(x,t)$. Similarly, the spatially-discretized
collection of white noise processes $\left(\D x\right)^{-1/2}\,\M{\mathcal{W}}$
can formally be associated with the space-time white noise $\M{\mathcal{Z}}$.

We take the coarse-grained Hamiltonian function to be the natural
(local equilibrium \cite{DiscreteDiffusion_Espanol}) discretization
of (\ref{eq:H_Burgers_continuum}), 
\begin{equation}
H\left(\V u\right)=\sum_{j=1}^{N}\frac{\D x}{2}u_{j}^{2},\label{eq:H_Burgers_discrete}
\end{equation}
We will construct a spatial discretization that leads to a finite-dimensional
generic Langevin equation of the form (\ref{eq:x_t_general})
\begin{equation}
\partial_{t}\V u=\M S\frac{\partial H}{\partial\V u}+\frac{\nu}{\D x}\M D_{2}\frac{\partial H}{\partial\V u}+\left(\frac{2\nu}{\D x}\right)^{1/2}\M D_{1}\M{\mathcal{W}}(t).\label{eq:u_t_Burgers}
\end{equation}
Here $\M{\mathcal{W}}$ is a vector of $N_{w}$ independent white-noise
processes (formally, time derivatives of independent Wiener processes),
$\M D_{1}$ is a matrix representing the spatial discretization of
the divergence operator, such that $\M D_{2}=-\M D_{1}\M D_{1}^{\star}$
is a symmetric negative-semidefinite discretization of the Laplacian
operator. This system of SODEs has as an invariant distribution the
Gibbs distribution (\ref{eq:GibbsBoltzmann}) if $\M S$ is an antisymmetric
matrix discretizing (\ref{eq:L_Burgers}) that satisfies
\begin{equation}
\left[\frac{\partial}{\partial\V u}\cdot\M S\left(\V u\right)\right]_{k}=\sum_{j}\frac{\partial S_{j,k}}{\partial u_{j}}=0\mbox{ for all }k.\label{eq:no_div_L}
\end{equation}
We now construct specific finite-difference operators for $\M D_{1}$
and $\M S$.

A particularly simple choice that also generalizes to higher dimensions
\cite{LLNS_S_k} is to associate fluxes with the half-grid points
(faces of the grid in higher dimensions), and to define
\[
\left(\M D_{1}\M{\mathcal{W}}\right)_{j}=\frac{\mathcal{W}_{j+\frac{1}{2}}-\mathcal{W}_{j-\frac{1}{2}}}{\D x},\mbox{ giving }\left(\M D_{1}^{\star}\M u\right)_{j+\frac{1}{2}}=-\frac{u_{j+1}-u_{j}}{\D x}.
\]
This construction gives the familiar three-point discrete Laplacian
($2d+1$ points in dimension $d$),
\begin{equation}
\left(\M D_{2}\M u\right)_{j}=\frac{u_{j-1}-2u_{j}+u_{j+1}}{\D x^{2}},\label{eq:DFDB_Burgers}
\end{equation}
and is therefore an attractive choice that satisfies the discrete
fluctuation-dissipation principle \cite{LLNS_S_k}. If periodic boundary
conditions are imposed, we set $u_{0}=u_{N}$ and $u_{N+1}=u_{1}$
and $\mathcal{W}_{\frac{1}{2}}=\mathcal{W}_{N+\frac{1}{2}}$ (i.e.,
$N_{w}=N$). For Dirichlet boundary conditions we fix $u_{0}$ and
$u_{N+1}$ at specified values and do not need to impose any boundary
conditions on $\mathcal{W}$ (i.e., $N_{w}=N+1$).

A natural choice for $\M S$ is formed by choosing a skew-adjoint
discretization $\widetilde{\M D}_{1}=-\widetilde{\M D}_{1}^{\star}$
of $\partial_{x}$, in general different from $\M D_{1}$, and discretizing
(\ref{eq:L_Burgers}) directly as 
\[
\left(\M S\V u\right)_{j}=-\frac{c}{3}\left[u_{j}\left(\widetilde{\M D}_{1}u\right)_{j}-\left(\widetilde{\M D}_{1}^{\star}u^{2}\right)_{j}\right]=-\frac{c}{3}\left[u_{j}\left(\widetilde{\M D}_{1}u\right)_{j}+\left(\widetilde{\M D}_{1}u^{2}\right)_{j}\right],
\]
where $u^{2}=\left\{ u_{1}^{2},\dots,u_{N}^{2}\right\} $. We choose
$\widetilde{\M D}_{1}$ to be the second-order centered difference
operator
\[
\left(\widetilde{\M D}_{1}\V u\right)_{j}=\frac{u_{j+1}-u_{j-1}}{2\D x},
\]
leading to an explicit expression that makes it clear that $\M S\V u$
is a discretization of $-cuu_{x}$,
\[
\left(\M S\V u\right)_{j}=-\frac{c}{3}\left[u_{j}\left(\frac{u_{j+1}-u_{j-1}}{2\D x}\right)+\frac{u_{j+1}^{2}-u_{j-1}^{2}}{2\D x}\right]=-c\left(\frac{u_{j-1}+u_{j}+u_{j+1}}{3}\right)\left(\frac{u_{j+1}-u_{j-1}}{2\D x}\right).
\]
The above discretization of the advective term has been considered
frequently in the literature, as discussed in detail in Ref. \cite{Hamiltonian_BurgersHopf}.
It can be seen as a weighted combination of the ``convective'' and
the ``conservative'' forms of advection \cite{ConservativeDifferences_Incompressible}
with weights $1/3$ and $2/3$, which is the unique choice of weights
that gives a conservative and skew-adjoint discretization of advection.
It is important to note that one can write the nonlinear term in conservative
form,
\begin{equation}
\left(\M S\V u\right)_{j}=-\frac{c}{2}\left(\frac{u_{j+\frac{1}{2}}^{2}-u_{j-\frac{1}{2}}^{2}}{\D x}\right),\mbox{ where }u_{j+\frac{1}{2}}^{2}=\frac{u_{j}^{2}+u_{j}u_{j+1}+u_{j+1}^{2}}{3}.\label{eq:advection_Burgers}
\end{equation}
Due to the skew-symmetry, in the absence of viscosity the total ``energy''
(\ref{eq:H_Burgers_discrete}) is conserved for periodic systems.
It can also easily be shown that that the condition (\ref{eq:div_free_L})
is satisfied and therefore this particular discretization of the advective
term preserves the Hamiltonian structure of the equations \cite{Hamiltonian_BurgersHopf}.

Putting the pieces together we can write the semi-discrete fluctuating
Burgers equation as a system of SODEs, $j=1,\dots,N$,
\begin{eqnarray}
\frac{du_{j}}{dt} & = & -\frac{c}{6\D x}\left(u_{j-1}+u_{j}+u_{j+1}\right)\left(u_{j+1}-u_{j-1}\right)\label{eq:semidiscrete_Burgers}\\
 & + & \frac{\nu}{\D x^{2}}\left(u_{j-1}-2u_{j}+u_{j+1}\right)+\frac{\left(2\nu\right)^{1/2}}{\D x^{3/2}}\left(\mathcal{W}_{j+\frac{1}{2}}(t)-\mathcal{W}_{j-\frac{1}{2}}(t)\right).\nonumber 
\end{eqnarray}
With periodic boundary conditions, this stochastic method of lines
\cite{FD_Nematics_Adhikari} discretization strictly conserves the
total energy (\ref{eq:H_Burgers_discrete}) and the total momentum
\[
m(\V u)=\sum_{j=1}^{N}\D x\, u_{j}.
\]
The equilibrium distribution is the discrete Gibbs-Boltzmann distribution
\[
P_{\text{eq}}\left(\V u\right)=Z^{-1}\exp\left[-\frac{\D x}{2}\sum_{j=1}^{N}u_{j}^{2}\right]\delta\left(\D x\sum_{j=1}^{N}u_{j}-m_{0}\right),
\]
where $m_{0}$ is the initial value for the total momentum. In the
next section, we construct efficient temporal discretizations of (\ref{eq:semidiscrete_Burgers})
that preserve these properties as well as possible.

\section{\label{sec:Weak-Accuracy}Weakly-Accurate Temporal Integrators}

In this section, we consider weak temporal integrators for coupled
systems of stochastic ordinary differential equations (SODEs) that
arise after spatial discretization of the equations of fluctuating
hydrodynamics, such as the system (\ref{eq:semidiscrete_Burgers}).
To make the discussion more general and applicable to a large range
of generic Langevin equations, we consider the system of nonlinear
additive-noise differential equations
\begin{equation}
\frac{d\V x}{dt}=\V a\left(\V x\right)+\M K\V{\mathcal{W}}\left(t\right)=\left[\V L\left(\V x\right)\right]\V x+\V g\left(\V x\right)+\M K\V{\mathcal{W}}\left(t\right).\label{eq:dx_dt_general}
\end{equation}
There are $N_{v}$ independent variables $\V x\left(t\right)$, and
$\V{\mathcal{W}}\left(t\right)$ denotes a collection of $N_{w}$
independent white-noise processes, formally identified with the time
derivative of a collection of independent Brownian motions (Wiener
processes). Here $\M K$ is a constant matrix, and we have used the
more natural differential notation since there is no difference between
the different stochastic interpretations (e.g., Ito and Stratonovich).
For fluctuating hydrodynamics applications, we split the drift $\V a\left(\V x\right)$
into a diffusive term $\left[\V L\left(\V x\right)\right]\V x$ and
an advective term $\V g\left(\V x\right)$. This becomes particularly
important when considering semi-implicit temporal discretizations
since diffusion often needs to be treated implicitly for stability
reasons. In general, $\V L\left(\V x\right)$ may depend on $\V x$
since the transport coefficients (e.g., viscosity) may depend on certain
state variables (e.g., density). 

Multiplicative noise poses well-known difficulties with constructing
higher-order temporal integrators \cite{WeakTrapezoidal}. There are
many important fluctuating hydrodynamic equations with additive noise,
such as the fluctuating Navier-Stokes equation (\ref{eq:LLNS_incomp_v}),
in which $\V L\left(\V x\right)\equiv\M L=\text{const}$ is a multiple
of the (discrete) Laplacian and $\V K\left(\V x\right)\equiv\M K=\text{const}$.
is a multiple of the (discrete) divergence. For generality in Appendix
\ref{Appendix-Accuracy} we also consider constructing first-order
accurate schemes for the case of multiplicative noise, where $\V K\left(\V x\right)$
also depends on $\V x$, and here we focus on second-order accuracy
for additive noise.

There is an extensive literature on numerical methods for finite-dimensional
systems of SDEs. At the same time, efficient numerical solution of
the types of systems of SDEs appearing in the stochastic method of
lines \cite{FD_Nematics_Adhikari} for fluctuating hydrodynamic equations
limits the choices of practicable techniques. One of the most important
characteristics of systems such as (\ref{eq:semidiscrete_Burgers})
is the presence of a large number of length scales and associated
relaxation times (hydrodynamic modes). This intrinsic stiffness is
particularly prominent for diffusive terms and makes the construction
of stochastic integrators particularly challenging. A powerful class
of integrators for systems with a broad spectrum of relevant time
scales are exponential integrators \cite{SDEWeakExponential,StochasticExponential}.
These integrators apply a local linearization \cite{LocalLinearization_SDE}
and use the matrix exponential $\exp\left(\M L\D t\right)$ to capture
the dynamics at a time step $\D t$ that under-resolves the stiffest
modes. Unfortunately, the computation of the matrix exponential is
only practicable when it is simple to diagonalize $\M L$, as is the
case when a Fourier basis is used for periodic systems \cite{StochasticImmersedBoundary}.
Multi-step schemes are widely used in the deterministic context because
of their favorable stability properties and reduced computational
effort per time step \cite{SDEMultistep_SmallNoise,CahnHilliard_WeakNoise}.
Weak multistep schemes have not received much attention in the literature,
and their analysis requires developing novel techniques that are outside
the scope of this work.

With these considerations in mind, we focus here on one-step Runge-Kutta
schemes. They have the advantage of potentially attaining high order
of accuracy without requiring evaluation of derivatives of $\V a\left(\V x\right)$.
We will consider numerical schemes that only require the generation
of $N_{w}=O\left(N_{v}\right)$ Gaussian random variables and do not
require the solution of \emph{nonlinear} systems of equations. While
this limits the robustness of the schemes in the nonlinear setting,
it is the only type of method that is practicable for the sort of
large systems of SODEs that arise when discretizing hydrodynamic SPDEs.
First we discuss the simpler case of explicit stochastic Runge-Kutta
integrators, and then we consider stochastic variants of two-stage
implicit-explicit Runge-Kutta (IMEX-RK) \cite{IMEX_PDEs} integrators.

\subsection{\label{sub:Conditions-Weak}Conditions for Second-Order Weak Accuracy}

In this section we discuss second-order weakly accurate \emph{one-step}
temporal discretizations with a constant time step $\D t$, and denote
the numerical approximation $\V x^{n}\approx\V x\left(n\D t\right).$
The most fundamental temporal integrator for (\ref{eq:dx_dt_general})
is the weakly first-order accurate Euler-Maruyama scheme,
\begin{equation}
\V x^{n+1}=\V x^{n}+\D t\,\V a^{n}+\D t^{\frac{1}{2}}\M K^{n}\V W^{n},\label{eq:Euler_Maryama}
\end{equation}
where the superscript denotes the time level at which the term is
evaluated, for example, $\V a^{n}\equiv\V a\left(\V x^{n}\right)$.
Here $\V W^{n}$ is a vector of $m$ independent standard Gaussian
variates (i.e., normally-distributed pseudorandom numbers with mean
zero and unit variance), generated independently at each time step.
The SDE (\ref{eq:dx_dt_general}) can, in fact, be defined through
the limit $\D t\rightarrow0$ of the scheme (\ref{eq:Euler_Maryama}).
The stochastic term $\D t^{\frac{1}{2}}\V W^{n}$ represents the (Wiener)
increment of the underlying Brownian motions over the time step. Alternatively,
one can view (\ref{eq:Euler_Maryama}) as an application of the deterministic
explicit Euler method to (\ref{eq:dx_dt_general}), with the \emph{discrete
white noise} $\D t^{-1/2}\V W^{n}$ representing the rough forcing
$\V{\mathcal{W}}\left(t\right)$. This viewpoint is particularly useful
when extending higher-order standard deterministic schemes to the
stochastic context.

Let us consider one-step schemes for the general nonlinear additive-noise
system of SDEs (\ref{eq:dx_dt_general}). The general theory of weak
accuracy for stochastic integrators is well-established and reviewed,
for example, in Section 2.2 of \cite{MilsteinSDEBook}. The key result
is that, under certain assumptions, second-order weak accuracy is
achieved if the first $2\cdot2+1=5$ moments of the numerical increment
$\D{\V x}^{n}=\V x^{n+1}-\V x^{n}$ match the moments of the increment
$\V x\left(n\D t+\D t\right)-\V x\left(n\D t\right)$ to order $O\left(\D t^{2}\right)$.
The required moments can be obtained from the well-known weak expansion
\begin{equation}
x_{\alpha}(n\D t+\Delta t)=x_{\alpha}(n\D t)+\D t^{\frac{1}{2}}K_{\alpha\beta}W_{\beta}^{n}+\Delta t\, a_{\alpha}^{n}+\frac{1}{2}\left(\Delta t^{2}\, a_{\gamma}^{n}+\Delta t^{\frac{3}{2}}K_{\gamma\epsilon}W_{\epsilon}^{n}\right)\left(\partial_{\gamma}a_{\alpha}^{n}\right)+\frac{\Delta t^{2}}{4}K_{\gamma\epsilon}K_{\delta\epsilon}\left(\partial_{\gamma}\partial_{\delta}a_{\alpha}^{n}\right)+O(\Delta t^{5/2}),\label{eq:2nd_weak_derivatives}
\end{equation}
where a repeated index implies summation and the short-hand notation
$\partial_{\gamma}\equiv\partial/\partial x_{\gamma}$ is employed.
This expansion is not directly useful for numerical approximations
since it requires evaluating derivatives of the drift function. Instead,
we employ Runge-Kutta schemes and then ensure second-order weak accuracy
by matching the moments of the numerical increments to (\ref{eq:2nd_weak_derivatives}).
The details of these calculations are given in Appendix \ref{Appendix-Accuracy}.

A well-known weakly second-order accurate predictor-corrector scheme
that is consistent with the conditions derived in Appendix \ref{Appendix-Accuracy}
is a two-stage explicit trapezoidal method \cite{KineticStochasticIntegral_Ottinger,RandomlyForcedTurbulence,WeakTrapezoidal}.
In this scheme the first stage is an Euler-Maruyama predictor, and
the corrector stage is an explicit trapezoidal rule, 
\begin{eqnarray}
\tilde{\boldsymbol{x}}^{n+1} & = & \boldsymbol{x}^{n}+\Delta t\,\boldsymbol{a}^{n}+\D t^{\frac{1}{2}}\boldsymbol{K}\boldsymbol{W}^{n},\nonumber \\
\boldsymbol{x}^{n+1} & = & \boldsymbol{x}^{n}+\frac{\Delta t}{2}(\tilde{\boldsymbol{a}}^{n+1}+\boldsymbol{a}^{n})+\D t^{\frac{1}{2}}\boldsymbol{K}\boldsymbol{W}^{n}.\label{eq:trapezoidal_PC}
\end{eqnarray}
One can naively obtain this scheme from the classical deterministic
predictor-corrector algorithm by thinking of $\M F^{n}=\D t^{-1/2}\boldsymbol{K}\boldsymbol{W}^{n}$
as a constant applied forcing. The fully explicit nature of this method
severely restricts the time step due to stability limits arising from
the stiff diffusive terms in fluctuating hydrodynamics. We will consider
semi-implicit RK schemes in more detail in Section \ref{sec:IMEX}.

\subsection{Equilibrium Fluctuation Spectrum}

An important property of Langevin-type equations, including those
of fluctuating hydrodynamics, is the existence of a non-trivial stationary
distribution (invariant measure). It is important for numerical schemes
to have an equilibrium distribution that is in, some appropriate sense,
close to that of the stochastic differential equations.\textbf{ }A
recently proposed-approach \cite{MetropolizedSDEs} is to add a Metropolis-Hastings
acceptance-rejection rule to a classical integrator such as the Euler-Maruyama
scheme. This ``Metropolization'' ensures that the equilibrium distribution
of the numerical approximation is controlled; however, this is done
at the cost of reducing the temporal accuracy because of rejections.
It is therefore important to ensure that the non-Metropolized numerical
scheme produces a good approximation to the equilibrium distribution,
so that rejections are infrequent.

Mattingly et al. \cite{InvariantMeasureWeak} show that in some appropriate
metric the invariant measure (which is assumed to exist) of the numerical
scheme has the same order of accuracy as the weak order of accuracy
over finite time intervals. This only provides an asymptotic error
bound, however, and does not provide an estimate of the actual error.
By focusing on the linearized equations of fluctuating hydrodynamics
one can easily obtain explicit estimates for the invariant measure
of a given numerical scheme and thus understand the nature of discretization
errors in the long-time dynamics. This approach was used by some of
us in an earlier publication \cite{LLNS_S_k} to analyze and improve
explicit Runge-Kutta schemes for compressible fluctuating hydrodynamics.
Here we briefly review the main results and discuss some generalizations.

We consider the linear system of additive-noise SDEs
\begin{equation}
\frac{d\V x}{dt}=\M L\V x+\M K\mathcal{\V W}\left(t\right).\label{eq:linearized_eq}
\end{equation}
A general linear one-step temporal scheme for this equation has the
form
\[
\V x^{n+1}=\M Q\V x^{n}+\D t^{\frac{1}{2}}\M R\V W^{n},
\]
where $\M Q$ and $\M R$ are some iteration matrices. Since this
is a linear equation forced by a Gaussian process, the solution is
a Gaussian process. The equilibrium or steady-state covariance $\M C_{\D t}=\av{\V x^{n}\left(\V x^{n}\right)^{\star}}$
of this linear iteration is the solution of the linear system (see,
for example, the derivation in \cite{LLNS_S_k})
\begin{equation}
\M Q\M C_{\D t}\M Q^{\star}-\M C_{\D t}=-\D t\,\M R\M R^{\star}.\label{eq:C_dt_eq}
\end{equation}
In the limit $\D t\rightarrow0$ any consistent and stable numerical
scheme should give the correct equilibrium covariance $\M C=\av{\V x\left(t\right)\V x^{\star}\left(t\right)}$,
which is the solution to \cite{GardinerBook,AMR_ReactionDiffusion_Atzberger,LLNS_S_k}
\begin{equation}
\M L\M C+\M C\M L^{\star}=-\M K\M K^{\star}.\label{eq:C_eq}
\end{equation}

Equation (\ref{eq:linearized_eq}) can easily be solved explicitly
to obtain an \emph{exact} \emph{exponential integrator} for which
$\M Q=\exp\left(\M L\D t\right)$ follows from the deterministic variation-of-constants
formula. This exponential scheme will be an exact integrator for (\ref{eq:linearized_eq})
if and only if
\begin{equation}
\M R\M R^{\star}=\D t^{-1}\left[\M C-\M Q\M C\M Q^{\star}\right]=\D t^{-1}\left[\M C-\exp\left(\M L\D t\right)\M C\exp\left(\M L^{\star}\D t\right)\right].\label{eq:N_cov}
\end{equation}
In general, one cannot write an explicit solution to this equation
unless one can explicitly diagonalize $\M L$ and $\M C$ in some
basis.

\subsubsection{Implicit Midpoint Rule}

Runge-Kutta schemes approximate the matrix exponential $\exp\left(\M L\D t\right)$
with a polynomial (for fully explicit schemes) or a rational (for
semi-implicit schemes) approximation. An important example is provided
by the implicit midpoint (equivalently, trapezoidal) method (Crank-Nicolson
scheme) applied to the linear problem (\ref{eq:linearized_eq}),
\begin{equation}
\boldsymbol{x}^{n+1}=\boldsymbol{x}^{n}+\frac{\Delta t}{2}\M L(\boldsymbol{x}^{n}+\boldsymbol{x}^{n+1})+\D t^{\frac{1}{2}}\M K\boldsymbol{W}^{n}.\label{eq:Crank-Nicolson}
\end{equation}
In this scheme the iteration matrix $\M Q$ is a $1-1$ Pade approximation
of the matrix exponential, 
\begin{equation}
\M Q=\left(\M I-\frac{\M L\D t}{2}\right)^{-1}\left(\M I+\frac{\M L\D t}{2}\right)=\exp\left(\M L\D t\right)+O\left(\D t^{3}\right),\label{eq:RatExp_mid}
\end{equation}
and $\M R=\left(\M I-\M L\D t/2\right)^{-1}\M K$. It is not hard
to show that the implicit midpoint scheme leads to the correct equilibrium
covariance $\M C$ for \emph{any} time step size since
\[
\M R\M R^{\star}=\D t^{-1}\left[\M C-\M Q\M C\M Q^{\star}\right],
\]
as seen from a straightforward explicit calculation, 
\begin{eqnarray*}
 &  & \D t^{-1}\left[\M C-\M Q\M C\M Q^{\star}\right]\\
 & = & \D t^{-1}\left(\M I-\frac{\M L\D t}{2}\right)^{-1}\left[\left(\M I-\frac{\M L\D t}{2}\right)\M C\left(\M I-\frac{\M L^{\star}\D t}{2}\right)-\left(\M I+\frac{\M L\D t}{2}\right)\M C\left(\M I+\frac{\M L^{\star}\D t}{2}\right)\right]\left(\M I-\frac{\M L^{\star}\D t}{2}\right)^{-1}\\
 & = & \left(\M I-\frac{\M L\D t}{2}\right)^{-1}\left(-\M L\M C-\M C\M L^{\star}\right)\left(\M I-\frac{\M L^{\star}\D t}{2}\right)^{-1}=\left(\M I-\frac{\M L\D t}{2}\right)^{-1}\M K\M K^{\star}\left(\M I-\frac{\M L^{\star}\D t}{2}\right)^{-1}=\M R\M R^{\star}.
\end{eqnarray*}
An alternative derivation of the fact that (\ref{eq:Crank-Nicolson})
gives the correct steady-state covariance for any time step size $\D t$
can be found in the Appendix of Ref. \cite{LLNS_Staggered}. That
derivation is based on showing that the iteration (\ref{eq:Crank-Nicolson})
is a Metropolis-Hastings Monte Carlo algorithm to sample the invariant
distribution of (\ref{eq:linearized_eq}).

The implicit midpoint rule can easily be generalized to the nonlinear
system (\ref{eq:dx_dt_general}),
\[
\boldsymbol{x}^{n+1}=\boldsymbol{x}^{n}+\Delta t\,\M a\left(\frac{\boldsymbol{x}^{n+1}+\boldsymbol{x}^{n}}{2}\right)+\D t^{\frac{1}{2}}\boldsymbol{K}\boldsymbol{W}^{n},
\]
which can be shown to be weakly second-order accurate. This scheme
is a particularly good candidate for solving Langevin-type equations
because it is a time-reversible and quasi-symplectic integrator \cite{Hamiltonian_Leimkuhler}
that exactly conserves all quadratic invariants (e.g., a quadratic
Hamiltonian). However, it requires the solution of a nonlinear system
of equations at every time step. This nonlinear system of equations
may not have a unique solution and is in general too expensive to
solve for large-scale hydrodynamic calculations. In the special case
of the stochastic Burgers or Navier-Stokes equations, the only nonlinearity
in $\V a\left(\V x\right)$ comes from the advective term, which has
the special form (to within irrelevant constants) $\left[\M S\left(\V x\right)\right]\V x$
and can be linearized as $\left[\M S\left(\V x^{n+\frac{1}{2}}\right)\right]\V x$,
where $\V x^{n+\frac{1}{2}}$ is a mid-point estimate that has to
be obtained via a predictor stage. Such an approach gives a scheme
that only requires solving a linear systems in each time step, while
still preserving quadratic invariants (e.g., total kinetic energy).
It is, however, not a time-reversible scheme. Furthermore, solving
the non-symmetric systems that arise when advection is discretized
in a semi-implicit manner poses a significant linear algebra challenge,
especially when constraints such as incompressibility are included.
For this reason, in the next section we consider implicit-explicit
Runge-Kutta schemes in which only diffusive terms are, potentially,
discretized implicitly.

\subsection{\label{sub:RK3Scheme}Fully Explicit Runge-Kutta Scheme}

We now illustrate how the conditions derived in Appendix \ref{Appendix-Accuracy}
can be used in practice to construct a fully-explicit three-stage
Runge-Kutta (RK3) integrator. In Ref. \cite{Bell:07}, an algorithm
for the solution of the compressible equations of fluctuating hydrodynamics
was proposed, which is based on a well-known three-stage, low-storage
total variation diminishing Runge-Kutta (RK3) scheme \cite{Gottlieb:98}.
The RK3 scheme is a simple discretization for the deterministic compressible
Navier-Stokes equations that is stable in the inviscid limit, even
when slope-limiters are omitted for the convective terms. A stochastic
version of the RK3 scheme was analyzed in Ref. \cite{LLNS_S_k} based
on a linearized analysis, and employed in Ref. \cite{LLNS_Staggered}
along with a staggered spatial discretization. In Ref. \cite{LLNS_S_k},
nonlinearities were not considered and therefore the scheme presented
there is not second-order weakly accurate for nonlinear equations. 

Each time step of the RK3 algorithm is composed of three stages, the
first one estimating $\V x$ at time $t=(n+1)\D t$, the second at
$t=(n+\frac{1}{2})\D t$, and the final stage obtaining a third-order
accurate estimate at $t=(n+1)\D t$. Each stage consists of an Euler-Maryama
step followed by weighted averaging with the value from the previous
stage, 
\begin{align}
\tilde{\boldsymbol{x}}^{n+1}= & \boldsymbol{x}^{n}+\Delta t\,\boldsymbol{a}^{n}+\D t^{\frac{1}{2}}\boldsymbol{K}\left(\alpha_{1}\boldsymbol{W}_{A}^{n}+\beta_{1}\boldsymbol{W}_{B}^{n}\right)\nonumber \\
\widetilde{\V x}^{n+\frac{1}{2}}= & \frac{3}{4}\V x^{n}+\frac{1}{4}\left[\widetilde{\V x}^{n+1}+\Delta t\,\tilde{\boldsymbol{a}}^{n+1}+\D t^{\frac{1}{2}}\boldsymbol{K}\left(\alpha_{2}\boldsymbol{W}_{A}^{n}+\beta_{2}\boldsymbol{W}_{B}^{n}\right)\right]\nonumber \\
\V x^{n+1}= & \frac{1}{3}\V x^{n}+\frac{2}{3}\left[\widetilde{\V x}^{n+\frac{1}{2}}+\Delta t\,\tilde{\boldsymbol{a}}^{n+\frac{1}{2}}+\D t^{\frac{1}{2}}\boldsymbol{K}\left(\alpha_{3}\boldsymbol{W}_{A}^{n}+\beta_{3}\boldsymbol{W}_{B}^{n}\right)\right].\label{RK3_explicit_stages}
\end{align}
Here $\V W_{A}^{n}$ and $\V W_{B}^{n}$ are two independent vectors
of i.i.d. normal random variates that are generated independently
at each RK3 step, and the weights $\V{\alpha}$ and $\V{\beta}$ are
to be determined. In principle one could use a third sample $\V W_{C}^{n}$;
however, this increases the cost of the method and is insufficient
to yield a weakly third-order accurate scheme. Following Ref. \cite{RK3_WeakAdditive},
we can fix $\alpha_{3}$ and $\beta_{3}$ by making the arbitrary
choice that after all the stages are combined the stochastic increment
in $\V x^{n+1}$ be $\D t^{\frac{1}{2}}\boldsymbol{K}\boldsymbol{W}_{A}^{n}$.

Runge-Kutta schemes of the above form have been analyzed in Ref. \cite{RK3_WeakAdditive}
and the moment conditions for weak accuracy of any order derived.
Up to second-order one can easily obtain these conditions by explicit
Taylor series expansion and comparison of the moments of the numerical
increment to those in (\ref{eq:2nd_weak_derivatives}). This gives
two quadratic equations for the weights $\V{\alpha}$ and $\V{\beta}$.
Two more equations can be obtained by asking for third-order accuracy
of the static covariance $\M C_{\D t}=\M C+\D t^{3}\D{\M C}+O(\D t^{4})$
in the linear case. This simple calculation consists of applying the
RK3 scheme to the linear equation (\ref{eq:linearized_eq}) in order
to extract an explicit expression for $\M Q$ and $\M R$, and then
substituting these expressions in (\ref{eq:C_dt_eq}) and using (\ref{eq:C_eq})
to eliminate $\M K\M K^{\star}$. By equating the coefficients in
front of terms of lower order in $\D t$, we can ensure that the error
in the stationary covariance is of order $\D t^{3}$. In particular,
equating the coefficients in front of the terms involving $\M L^{3}\M C$
and $\M L^{2}\M C\M L^{\star}$ to zero gives two additional quadratic
equations for the weights $\V{\alpha}$ and $\V{\beta}$. The solution
of the resulting system of four equations for the weights $\alpha_{1},\,\alpha_{2,\,}\beta_{1}$
and $\beta_{2}$ gives a stochastic RK3 scheme that is of weak order
two in the general case and of order three in the linear case,
\begin{align*}
\alpha_{1}= & \alpha_{2}=\alpha_{3}=1\\
\beta_{1}= & \frac{\left(2\,\sqrt{2}\pm\sqrt{3}\right)}{5}\\
\beta_{2}= & \frac{\left(-4\,\sqrt{2}\pm3\,\sqrt{3}\right)}{5}\\
\beta_{3}= & \frac{\left[\sqrt{2}\mp2\sqrt{3}\right]}{10}.
\end{align*}
For fluctuating hydrodynamics applications, we recommend the upper
sign since it gives better discrete static structure factors for a
model one-dimensional stochastic advection-diffusion equation (see
Ref. \cite{LLNS_S_k} for an illustration of this type of calculations).

The RK3 scheme (\ref{RK3_explicit_stages}) suffers from a severe
time step limitation due to the explicit handling of the diffusive
terms. We consider alternative semi-implicit methods next.

\subsection{\label{sec:IMEX}Semi-Implicit Runge-Kutta Temporal Integrators}

In this section, we construct two-stage second-order implicit-explicit
Runge-Kutta schemes for solving a system of SDEs 
\[
\frac{d\V x}{dt}=\boldsymbol{L}(\boldsymbol{x})\boldsymbol{x}+\boldsymbol{g}(\boldsymbol{x})+\boldsymbol{K}\boldsymbol{\mathcal{W}}\left(t\right),
\]
where $\V g\left(\V x\right)$ denotes all of the terms handled explicitly
(e.g., advection or external forcing). We will consider schemes that
at time step $n$ require solving only linear systems involving the
matrix $\boldsymbol{L}^{n}=\boldsymbol{L}(\boldsymbol{x}^{n}),$ and
are parametrized by a vector of weights $\V w$ (Butcher tableau).
The first stage in any of these schemes is a predictor step to estimate
$\tilde{\boldsymbol{x}}\approx\V x\left(n\D t+w_{2}\D t\right)$,
where $w_{2}$ is some chosen weight (e.g., $w_{2}=1/2$ for a midpoint
predictor). The corrector completes the step by estimating $\boldsymbol{x}^{n+1}$
at time $(n+1)\D t$, 
\begin{eqnarray}
\tilde{\boldsymbol{x}} & = & \boldsymbol{x}^{n}+(w_{2}-w_{1})\Delta t\,\boldsymbol{L}^{n}\boldsymbol{x}^{n}+w_{1}\Delta t\,\boldsymbol{L}^{n}\tilde{\boldsymbol{x}}+w_{2}\Delta t\,\boldsymbol{g}^{n}+\left(w_{2}\Delta t\right)^{\frac{1}{2}}\boldsymbol{K}\boldsymbol{W}_{1}^{n},\nonumber \\
\boldsymbol{x}^{n+1} & = & \boldsymbol{x}^{n}+(1-w_{3}-w_{4})\Delta t\,\boldsymbol{L}^{n}\boldsymbol{x}^{n}+w_{3}\Delta t\,\boldsymbol{L}^{n}\tilde{\boldsymbol{x}}+w_{4}\Delta t\,\boldsymbol{L}^{n}\boldsymbol{x}^{n+1}\nonumber \\
 & + & w_{5}\Delta t\,(\widetilde{\boldsymbol{L}}-\boldsymbol{L}^{n})\tilde{\boldsymbol{x}}+w_{5}\Delta t\,\tilde{\boldsymbol{g}}+(1-w_{5})\Delta t\,\boldsymbol{g}^{n}\nonumber \\
 & + & \left(w_{2}\Delta t\right)^{\frac{1}{2}}\boldsymbol{K}\boldsymbol{W}_{1}^{n}+\left(\left(1-w_{2}\right)\Delta t\right)^{\frac{1}{2}}\boldsymbol{K}\boldsymbol{W}_{2}^{n}.\label{eq:generic_RK2}
\end{eqnarray}
The intuition behind the handling of the stochastic increments in
the predictor/corrector stages is that $\left(w_{2}\Delta t\right)^{1/2}\boldsymbol{W}_{1}^{n}$
samples the increment of the underlying Wiener processes over the
time interval $w_{2}\D t$, while $\left(\left(1-w_{2}\right)\Delta t\right)^{1/2}\boldsymbol{W}_{2}^{n}$
represents the independent increment over the remainder of the time
step.

According to the calculations detailed in Appendix \ref{Appendix-Accuracy},
in order to be second-order weakly accurate in the case of additive
noise, the weights $\V w$ should satisfy the conditions 
\begin{equation}
w_{2}w_{5}=\frac{1}{2},\quad w_{2}w_{3}+w_{4}=\frac{1}{2}.\label{eq:conditions_2nd}
\end{equation}
We now catalog some possible choices of these weights that satisfy
these conditions. We will test the performance of these schemes on
the fluctuating Burgers and Navier-Stokes equations in Section \ref{sec:Numerical-Results}.

\subsubsection{\label{sub:Explicit-Midpoint}Explicit Midpoint Predictor-Corrector
Scheme}

If $w_{1}=0$ and $w_{4}=0$ we obtain fully explicit schemes. In
this case $w_{3}=w_{5}$ and therefore the splitting into implicit
and explicit parts does not matter, and the only function that appears
in the scheme is $\V a(\V x)=\boldsymbol{L}(\boldsymbol{x})\boldsymbol{x}+\boldsymbol{g}(\boldsymbol{x})$.
The only parameter of choice is $w_{2}$. A well-known method that
fits the above format is the explicit trapezoidal predictor-corrector
method, $w_{2}=1$, and therefore $w_{5}=w_{3}=\frac{1}{2}$. This
is exactly the scheme (\ref{eq:trapezoidal_PC}). For linear equations,
this scheme gives static covariances accurate to second-order, $\M C_{\D t}=\M C+\D t^{2}\D{\M C}+O(\D t^{3})$.
A notable advantage of this scheme is that it requires generating
only a single random increment per time step.

In order to choose the ``best'' $w_{2}$, we can look at the accuracy
of the static covariance in the linear case, just as we did for the
RK3 scheme in Section \ref{sub:RK3Scheme}. A simple calculation shows
that in order to obtain third-order accuracy of the static covariance,
$\M C_{\D t}=\M C+\D t^{3}\D{\M C}+O(\D t^{4})$, we need to use a
mid-point predictor stage, $w_{2}=1/2$ and therefore $w_{5}=w_{3}=1$.
This gives the explicit midpoint predictor-corrector method
\begin{eqnarray}
\tilde{\boldsymbol{x}}^{n+\frac{1}{2}} & = & \boldsymbol{x}^{n}+\frac{\Delta t}{2}\,\boldsymbol{a}^{n}+\left(\frac{\Delta t}{2}\right)^{\frac{1}{2}}\boldsymbol{K}\boldsymbol{W}_{1}^{n}\nonumber \\
\boldsymbol{x}^{n+1} & = & \boldsymbol{x}^{n}+\D t\,\tilde{\boldsymbol{a}}^{n+\frac{1}{2}}+\left(\frac{\Delta t}{2}\right)^{\frac{1}{2}}\boldsymbol{K}\left(\boldsymbol{W}_{1}^{n}+\boldsymbol{W}_{2}^{n}\right),\label{eq:midpoint_PC}
\end{eqnarray}
which requires generating two random increments per time step. It
is important to point out that for advection-diffusion problems the
time step $\D t$ needs to be substantially smaller than the stability
limit for the explicit midpoint scheme (\ref{eq:midpoint_PC}) to
realize its asymptotic order of accuracy and give substantially more
accurate static covariances than the explicit trapezoidal scheme (\ref{eq:trapezoidal_PC})
\cite{LLNS_S_k}.

\subsubsection{Implicit Trapezoidal Predictor-Corrector Scheme}

We now extend the implicit midpoint scheme (\ref{eq:Crank-Nicolson})
to the case when some terms, such as diffusion, are handled implicitly.
In Ref. \cite{LLNS_Staggered} the fluctuating Navier-Stokes equation
was solved using a scheme for which $w_{1}=w_{4}=w_{5}=1/2,\, w_{2}=1,\, w_{3}=0$,
giving the implicit trapezoidal predictor-corrector method, 
\begin{eqnarray}
\tilde{\boldsymbol{x}}^{n+1} & = & \boldsymbol{x}^{n}+\frac{\Delta t}{2}\boldsymbol{L}^{n}\left(\boldsymbol{x}^{n}+\tilde{\boldsymbol{x}}^{n+1}\right)+\Delta t\,\boldsymbol{g}^{n}+\D t^{\frac{1}{2}}\boldsymbol{K}\boldsymbol{W}^{n}\nonumber \\
\boldsymbol{x}^{n+1} & = & \boldsymbol{x}^{n}+\frac{\Delta t}{2}\boldsymbol{L}^{n}\left(\boldsymbol{x}^{n}+\boldsymbol{x}^{n+1}\right)+\frac{\Delta t}{2}(\widetilde{\boldsymbol{L}}^{n+1}-\boldsymbol{L}^{n})\tilde{\boldsymbol{x}}^{n+1}+\frac{\Delta t}{2}(\tilde{\boldsymbol{g}}^{n+1}+\boldsymbol{g}^{n})+\D t^{\frac{1}{2}}\boldsymbol{K}\boldsymbol{W}^{n}.\label{eq:impl_trapezoidal}
\end{eqnarray}
If $\V g=\V 0$ this scheme is stable for any time step, more precisely,
it is $A$-stable (see Appendix \ref{sub:L0-stable} for $L$-stable
schemes). The scheme has the great advantage that for the linearized
equation (\ref{eq:linearized_eq}) it gives the correct stationary
covariance regardless of the time step, while only requiring a single
random increment per time step. We emphasize that the \emph{dynamics}
of the fluctuations is \emph{not} correctly reproduced for large time
step sizes, as we discuss further in Appendix \ref{sub:L0-stable}.

\subsubsection{Implicit Midpoint Predictor-Corrector Scheme}

In the scheme (\ref{eq:impl_trapezoidal}), a trapezoidal approximation
is used for the explicit fluxes, just as in (\ref{eq:trapezoidal_PC}).
Another candidate is to use a midpoint approximation for the explicit
fluxes, as in (\ref{eq:midpoint_PC}), obtained by using $w_{2}=1/2$
and therefore $w_{5}=1$. We also choose a corrector stage where the
implicit part is the implicit midpoint rule (\ref{eq:Crank-Nicolson}),
which requires choosing $w_{3}=0$ and $w_{4}=1/2$. We are left with
a choice for $w_{1}$ in the predictor stage, 
\begin{eqnarray}
\tilde{\V x}^{n+\frac{1}{2}} & = & \boldsymbol{x}^{n}+\left(\frac{1}{2}-w_{1}\right)\boldsymbol{L}^{n}\boldsymbol{x}^{n}\Delta t+w_{1}\boldsymbol{L}^{n}\tilde{\V x}^{n+\frac{1}{2}}+\frac{\Delta t}{2}\boldsymbol{g}^{n}+\left(\frac{\Delta t}{2}\right)^{\frac{1}{2}}\boldsymbol{K}\boldsymbol{W}_{1}^{n}\nonumber \\
\boldsymbol{x}^{n+1} & = & \boldsymbol{x}^{n}+\frac{\Delta t}{2}\boldsymbol{L}^{n}\left(\boldsymbol{x}^{n}+\boldsymbol{x}^{n+1}\right)+\Delta t(\widetilde{\boldsymbol{L}}^{n+\frac{1}{2}}-\boldsymbol{L}^{n})\tilde{\boldsymbol{x}}^{n+\frac{1}{2}}+\Delta t\,\tilde{\V g}^{n+\frac{1}{2}}+\left(\frac{\Delta t}{2}\right)^{\frac{1}{2}}\boldsymbol{K}\left(\boldsymbol{W}_{1}^{n}+\boldsymbol{W}_{2}^{n}\right).\label{eq:impl_midpoint}
\end{eqnarray}
Two obvious choices are $w_{1}=1/2$, which for linear equations makes
the predictor stage a backward Euler step with time step size $\D t/2$.
An alternative is to use $w_{1}=1/4$, which for linear equations
makes the predictor stage an implicit midpoint step with time step
size $\D t/2$. For the linearized equation (\ref{eq:linearized_eq})
the predictor stage does not actually matter since it is only used
in evaluating the nonlinear terms in the corrector stage. Therefore,
we will compare the implicit midpoint schemes with $w_{1}=1/4$ and
$w_{1}=1/2$ numerically in Section \ref{sec:Numerical-Results}.

\section{\label{sec:FluctNS}Fluctuating Navier-Stokes Equation}

The implicit-explicit schemes discussed in Section \ref{sec:IMEX}
are general schemes suitable for unconstrained SDEs and cannot directly
be applied to the fluctuating Navier-Stokes equation (\ref{eq:LLNS_incomp_v}).
Some care is required in handling the incompressibility constraint
in a computationally-efficient manner without compromising the stochastic
accuracy. The spatio-temporal discretization we analyze here was proposed
and applied in Ref. \cite{LLNS_Staggered}; here we provide additional
analysis and a discussion of alternative approaches.

\subsection{Continuum Equations}

In principle, the incompressibility constraint can be most easily
handled by using a projection operator formalism to eliminate pressure
from (\ref{eq:LLNS_incomp_v}) and write the fluctuating Navier-Stokes
equation in the form
\begin{equation}
\partial_{t}\V v=\M{\mathcal{P}}\left[-\V v\cdot\grad\V v+\nu\grad^{2}\V v+\left(2\nu\rho^{-1}\, k_{B}T\right)^{\frac{1}{2}}\grad\cdot\M{\mathcal{Z}}_{\V v}\right].\label{eq:fluct_NS_P}
\end{equation}
Here $\M{\mathcal{P}}$ is the orthogonal projection onto the space
of divergence-free velocity fields, $\M{\mathcal{P}}=\M I-\M{\mathcal{G}}\left(\M{\mathcal{D}}\M{\mathcal{G}}\right)^{-1}\M{\mathcal{D}}$
in real space, where $\M{\mathcal{D}}\equiv\grad\cdot$ denotes the
divergence operator and $\M{\mathcal{G}}\equiv\grad$ the gradient
operator with the appropriate boundary conditions taken into account.
We only consider periodic, no-slip and free-slip boundaries. With
periodic boundaries we can express all operators in Fourier space
and $\widehat{\M{\mathcal{P}}}=\M I-k^{-2}(\V k\V k^{\star})$, where
$\V k$ is the wavenumber. The application of the projection to the
right hand side ensures that $\grad\cdot\V v=0$ at all times if the
initial condition is divergence free. The divergence-free constraint
is a constant linear constraint and the projection restricts the velocity
dynamics to the constant linear subspace of divergence-free vector
fields. The projection operator can be applied in more general settings,
notably, in cases where the constraints are nonlinear and the noise
is multiplicative; however, the resulting expressions are rather complex
especially in the stochastic setting \cite{ProjectionConstrainedLangevin,ConstrainedStochasticDiffusion}.

In practice, the fluctuating velocities modeled by (\ref{eq:fluct_NS_P})
advect other quantities, and it is this coupling between the velocity
and other equations that is of most interest. Perhaps the simplest
example is provided by a stochastic advection-diffusion for the concentration
or density $c\left(\V r,t\right)$ of a large collection of non-interacting
\emph{passive tracers}. For example, $c\left(\V r,t\right)$ might
corresponds to the light intensity pattern of fluorescently-labeled
molecules suspended in the fluid in a Fluorescence Recovery After
Photobleaching (FRAP) experiment. In general, the equation for the
concentration has multiplicative noise \cite{LLNS_Staggered}. This
arises because the coarse-grained free energy functional $H\left[\V x\left(\V r,t\right)\right]=H\left(\V v,\, c\right)$
includes a contribution from the entropy of the passive tracer which
is, in general, a non-quadratic even if local functional of $c$.
For illustration purposes we can take a separable quadratic Hamiltonian
(i.e., independent Gaussian fluctuations in velocity and concentration),
\[
H\left(\V v,\, c\right)=H_{\V v}\left(\V v\right)+H_{c}\left(c\right)=\frac{\rho}{2}\int v^{2}\, d\V r+\frac{k_{B}T}{2\epsilon}\int c^{2}\, d\V r,
\]
and write the the model additive-noise tracer equation
\begin{equation}
\partial_{t}c=-\V v\cdot\grad c+\chi\grad^{2}c+\grad\cdot\left[\left(2\epsilon\chi\right)^{\frac{1}{2}}\M{\mathcal{Z}}_{c}\right].\label{eq:fluct_c_eq}
\end{equation}
The multiplicative noise case is not considered herein. Physically,
$\epsilon$ measures the degree of coarse graining, $\epsilon\sim N_{p}^{-1}$,
where $N_{p}$ is the number of tracer particles per coarse degree
of freedom. Note that (\ref{eq:fluct_c_eq}) is a conservation law
because $\V v\cdot\grad c=\grad\cdot\left(c\V v\right)$ due to incompressibility.

The coupled velocity-concentration system (\ref{eq:fluct_NS_P},\ref{eq:fluct_c_eq})
can formally be written in the form (\ref{eq:x_t_general}). The chemical
potential $\mu\left(c\right)=\partial H/\partial c\sim c$. The mobility
operator can be written as a sum of a skew-adjoint and a self-adjoint
part,
\begin{equation}
\M N=\M M-\M S=-\left[\begin{array}{cc}
\rho^{-1}\nu\left(\M{\mathcal{P}}\grad^{2}\M{\mathcal{P}}\right) & \V 0\\
\V 0 & \epsilon\left(k_{B}T\right)^{-1}\left(\chi\grad^{2}\right)
\end{array}\right]-\rho^{-1}\left[\begin{array}{cc}
\left(\M{\mathcal{P}}\M{\omega}\M{\mathcal{P}}\right) & \M{\mathcal{P}}\grad c\\
-\left(\grad c\right)^{T}\M{\mathcal{P}} & 0
\end{array}\right],\label{eq:N_NS}
\end{equation}
where $\M{\omega}$ is the antisymmetric vorticity tensor, $\M{\omega}_{jk}=\partial v_{k}/\partial r_{j}-\partial v_{j}/\partial r_{k}$,
and we used the vector identity 
\[
\M{\omega}\M v=-\left(\grad\times\V v\right)\times\V v=-\V v\cdot\grad\V v+\grad\left(\frac{v^{2}}{2}\right).
\]
Even though by skew symmetry the top right sub-block of $\M S$ is
nonzero, there is no coupling of concentration back in the velocity
equation because
\[
\left(\frac{\partial H}{\partial c}\right)\grad c=\left(\frac{dH_{c}}{dc}\right)\grad c=\grad H_{c}
\]
is a gradient of a scalar and is eliminated by the projection. The
velocity equation therefore remains of the form (\ref{eq:fluct_NS_P}).

\subsection{Spatial Discretization}

For a detailed description of the spatial discretization of (\ref{eq:fluct_NS_P},\ref{eq:fluct_c_eq})
that we employ we refer the reader to Ref. \cite{LLNS_Staggered}.
The discretization of the velocity equation is based on a staggered
or MAC grid \cite{HarWel65} in which the component of velocity along
a given dimension is discretized on a uniform grid that is shifted
by half a grid spacing along that dimension. Following the stochastic
methods of lines that we used for the Burgers equation in Section
\ref{sub:BurgersFD}, the spatial discretization of (\ref{eq:LLNS_incomp_v})
leads to a system of SODEs of the form
\begin{equation}
\frac{d\V v}{dt}=\M{\Set P}\left[\M S_{\V v}\left(\V v\right)\V v+\nu\M L_{\V v}\V v+\left(\frac{2\nu k_{B}T}{\rho\D V}\right)^{\frac{1}{2}}\,\M D_{\V w}\V{\mathcal{W}}_{\V v}\left(t\right)\right],\label{eq:spatial_incomp}
\end{equation}
where $\M S_{\V v}\left(\V v\right)$ denotes a discretization of
the advective operator $-\left(\V v\cdot\grad\right)$, $\D V$ is
the volume of a hydrodynamic cell, and $\V{\mathcal{W}}_{\V v}\left(t\right)$
is a collection of white-noise processes \cite{LLNS_Staggered}. Here
$\M D_{\V w}$ a tensor divergence operator that applies the conservative
discrete vector divergence operator $\M D$ independently for each
coordinate, $\M L_{\V v}$ is a discrete (vector) Laplacian, and $\M{\Set P}=\M I-\M G\left(\M D\M G\right)^{-1}\M D$
is a \emph{discrete projection} operator, where $\M G$ is a discrete
scalar gradient operator. The imposed periodic, no-slip or free-slip
boundary conditions are encoded in the specific forms of the discrete
difference operators near the boundaries of the domain.

Let us first focus on creeping Stokes flow, where the advective term
$\V v\cdot\grad\V v$ is neglected. Following the same procedure as
we employed for the fluctuating Burgers equation in Section \ref{sub:BurgersFD},
the spatial discretization is constructed to obey a discrete fluctuation-dissipation
balance principle. This relies on several key properties of the staggered
difference operators. Importantly, the discrete gradient and divergence
operators obey the duality relation $\M G=-\M D^{\star}$, just as
the continuum operators. The resulting scalar Laplacian $\M L_{s}=\M D\M G=-\M D\M D^{\star}$
is the standard ($2d+1$)-point discrete Laplacian, which is also
applied to each (staggered) component of the velocity to form $\M L_{\V v}=-\M D_{\V w}\left(\M D_{\V w}\right)^{\star}$
(see Ref. \cite{LLNS_Staggered} for a discussion of modifications
near physical boundaries). Because of the duality between $\M D$
and $\M G$ the MAC projection is self-adjoint, $\M{\Set P}^{\star}=\M{\Set P}$,
and idempotent, $\M{\Set P}^{2}=\M{\Set P}$, just like the continuum
projection operator. From these properties and Eq. (\ref{eq:C_eq})
it follows (see Appendix in Ref. \cite{LLNS_Staggered} for details)
that the equilibrium covariance of the fluctuating velocities is 
\begin{equation}
\left\langle \V v\V v^{\star}\right\rangle =\frac{k_{B}T}{\rho\D V}\,\M{\Set P}\label{eq:C_v_cont}
\end{equation}
This means that when an equilibrium snapshot of the velocity is expressed
in any orthonormal basis for the subspace of discretely divergence-free
vector fields, the coefficients are i.i.d. Gaussian random variables
with mean zero and variance $\rho^{-1}k_{B}T/\D V$. This is the expression
of discrete fluctuation-dissipation balance for the case of incompressible
flow.

The addition of the nonlinear advective term does not affect the discrete
fluctuation-dissipation balance since the advective term is skew-adjoint,
$\M S_{\V v}^{\star}=-\M S_{\V v}$, just like the discretization
(\ref{eq:advection_Burgers}) of the term $uu_{x}$ for the one-dimensional
case. Specifically, in two dimensions, for a given $\V u$ such that
$\M D\V u=0$, the spatial discretization of the advective term described
in Refs. \cite{ConservativeDifferences_Incompressible,LLNS_Staggered}
leads to
\begin{eqnarray*}
\left[\M S_{\V v}\left(\V u\right)\V v\right]_{i+\frac{1}{2},j}^{(x)} & = & -\left(4\D x\right)^{-1}\left[\left(u_{i+\frac{3}{2},j}^{(x)}+u_{i+\frac{1}{2},j}^{(x)}\right)v_{i+\frac{3}{2},j}^{(x)}-\left(u_{i-\frac{1}{2},j}^{(x)}+u_{i+\frac{1}{2},j}^{(x)}\right)v_{i-\frac{1}{2},j}^{(x)}\right]\\
 &  & -\left(4\D y\right)^{-1}\left[\left(u_{i,j+\frac{1}{2}}^{(y)}+u_{i+1,j+\frac{1}{2}}^{(y)}\right)v_{i+\frac{1}{2},j+1}^{(x)}-\left(u_{i,j-\frac{1}{2}}^{(y)}+u_{i+1,j-\frac{1}{2}}^{(y)}\right)v_{i+\frac{1}{2},j-1}^{(x)}\right].
\end{eqnarray*}
This can easily be shown to be a skew-adjoint discretization, $\left[\M S_{\V v}\left(\V u\right)\V v\right]\cdot\V w=-\left[\M S_{\V v}\left(\V u\right)\V w\right]\cdot\V v$,
for either periodic, free-slip and no-slip conditions (or any combination
thereof). Furthermore, this discretization leads to Hamiltonian dynamics
for inviscid flow, i.e., the phase-space flow generated by the advective
term is incompressible, 
\[
\frac{\partial}{\partial\V v}\cdot\M S_{\V v}\left(\V v\right)=\V 0.
\]
A good temporal integrator should preserve this special structure
of the equations and reproduce the correct velocity fluctuations for
reasonably large time step sizes.

Note that the addition of a passively-advected scalar field poses
no additional difficulties if $c$ is discretized on the regular (non-staggered)
grid underlying the (staggered) velocity grid. Specifically, the spatial
discretization of (\ref{eq:fluct_c_eq}) that we employ is a scalar
equivalent of (\ref{eq:spatial_incomp}),
\begin{equation}
\frac{d\V c}{dt}=\M S_{c}\left(\V v\right)\V c+\chi\M L_{c}\V c+\left(\frac{2\epsilon\chi}{\D V}\right)^{\frac{1}{2}}\M D\V{\mathcal{W}}_{c}\left(t\right),\label{eq:spatial_c_eq}
\end{equation}
where $\M S_{c}\left(\V v\right)$ is a cell-centered conservative
and skew-adjoint discretization of the advection operator $-\left(\V v\cdot\grad\right)$,
\[
\left[\M S_{c}\left(\V u\right)\V c\right]_{i,j}=-\left(2\D x\right)^{-1}\left(u_{i+\frac{1}{2},j}^{(x)}c_{i+1,j}-u_{i-\frac{1}{2},j}^{(x)}c_{i-1,j}\right)-\left(2\D y\right)^{-1}\left(u_{i,j+\frac{1}{2}}^{(y)}c_{i,j+1}-u_{i,j-\frac{1}{2}}^{(y)}c_{i,j-1}\right),
\]
and $\M L_{c}$ is equivalent to $\M L_{s}=\M D\M G$ except near
boundaries. The semi-discrete equation (\ref{eq:spatial_c_eq}) obeys
a discrete fluctuation-dissipation balance principle \cite{LLNS_Staggered}.
Specifically, at thermodynamic equilibrium the fluctuations in the
concentration are Gaussian with covariance 
\begin{equation}
\av{\V c\V c^{\star}}=\frac{\epsilon}{\D V}\M I,\label{eq:C_c_cont}
\end{equation}
and also uncorrelated with the velocity fluctuations.

The coupled velocity-concentration equation system of SODEs (\ref{eq:spatial_incomp},\ref{eq:spatial_c_eq})
can be written in the generic Langevin form (\ref{eq:x_t_general}).
Note that the concentration-dependent term in the velocity equation
that ought to be included to preserve the skew-symmetry of the non-diffusive
terms can be written in the form of a projected \emph{discrete} gradient
of a scalar, $\M{\Set P}\M G\left(c^{2}/2\right)$, which vanishes
identically. This shows that in two dimensions the system (\ref{eq:spatial_incomp},\ref{eq:spatial_c_eq})
is time reversible with respect to the equilibrium Gibbs-Boltzmann
distribution with the separable discrete Hamiltonian 
\[
H\left(\V v,c\right)=\frac{\rho\D V}{2}\sum_{i,j}\left[\left(v_{i+\frac{1}{2},j}^{(x)}\right)^{2}+\left(v_{i,j+\frac{1}{2}}^{(y)}\right)^{2}\right]+\frac{k_{B}T\D V}{2\epsilon}\sum_{i,j}c_{i,j}^{2}.
\]
Note that the Gibbs-Boltzmann distribution is constrained to the linear
subspace of discretely divergence-free vector fields, 
\[
\left(\M D\V v\right)_{i,j}=\D x^{-1}\left(v_{i+\frac{1}{2},j}^{(x)}-v_{i-\frac{1}{2},j}^{(x)}\right)+\D y^{-1}\left(v_{i,j+\frac{1}{2}}^{(y)}-v_{i,j-\frac{1}{2}}^{(y)}\right)=0,
\]
and with periodic boundaries both the average momentum and average
concentration are conserved by the dynamics.

\subsection{Temporal Discretization}

In our initial discussion of temporal integration schemes for (\ref{eq:spatial_incomp})
we will neglect the advective term and focus on creeping Stokes flow,
thus avoiding technical details while preserving the essential features
of the problem. There is a vast literature on deterministic temporal
integration of the incompressible Navier-Stokes equations, and, in
particular, the handling of the $\grad\pi$ term. One of the most
popular class of methods are splitting or \emph{projection methods},
such as the prototype projected Euler-Maruyama method,
\begin{equation}
\V v^{n+1}=\V v^{n}+\nu\D t\,\M{\Set P}\M L_{\V v}\V v^{n}+\left(2\nu\D t\right)^{\frac{1}{2}}\M{\Set P}\M D_{\V w}\V W_{\V v}^{n},\label{eq:projected_EM_exact}
\end{equation}
where we set $\rho^{-1}k_{B}T/\D V=1$ for simplicity and $\V W_{\V v}^{n}$
are i.i.d. standard normal random variates generated independently
at each time step. Note that in practice, due to roundoff errors and
the use of inexact Poisson solvers in the projection operation, it
is preferable to apply $\M{\Set P}$ to $\V v^{n}$ as well.

As explained in Appendix \ref{sec:ApproxProj} (see also Appendix
B in Ref. \cite{LLNS_Staggered}), the iteration (\ref{eq:projected_EM_exact})
gives a steady-state covariance that is a first-order accurate approximation
to the continuum result (\ref{eq:C_v_cont}),
\[
\M C_{\V v}=\av{\V v^{n}\left(\V v^{n}\right)^{\star}}=\M{\Set P}+\D t\,\D{\M C}_{\V v}+O\left(\D t^{2}\right).
\]
In Appendix \ref{sec:ApproxProj} we consider approximate projection
methods \cite{almgrenBellSzymczak:1996,ApproximateProjection_I} and
find that they do not satisfy this requirement. The scheme (\ref{eq:projected_EM_exact})
can be seen as a direct application of the Euler-Maruyama method to
(\ref{eq:spatial_incomp}). Note that any purely explicit scheme,
including the RK3 scheme described in Section \ref{sub:RK3Scheme},
can be applied to (\ref{eq:spatial_incomp}) by simply performing
a projection operation after every stage of the scheme.

Semi-implicit schemes can also be applied to (\ref{eq:spatial_incomp}).
As a prototype example, let us consider the implicit midpoint method
(\ref{eq:Crank-Nicolson}),
\begin{equation}
\V v^{n+1}=\V v^{n}+\frac{\nu\D t}{2}\,\M{\Set P}\M L_{\V v}\left(\V v^{n}+\V v^{n+1}\right)+\left(2\nu\D t\right)^{\frac{1}{2}}\M{\Set P}\M D_{\V w}\V W_{\V v}^{n}.\label{eq:incompressible_CN_P}
\end{equation}
At first sight, it appears that solving (\ref{eq:incompressible_CN_P})
requires the application of $\left[\M I-\left(\nu\D t/2\right)\M{\Set P}\M L_{\V v}\right]^{-1}$.
However, it is not hard to see that solving (\ref{eq:incompressible_CN_P})
is equivalent to solving the following linear system for the velocity
$\M v^{n+1}$ and the pressure $\V{\pi}^{n+\frac{1}{2}}$,
\begin{eqnarray}
\left(\M I-\frac{\nu\D t}{2}\M L_{\V v}\right)\V v^{n+1}+\D t\,\M G\V{\pi}^{n+\frac{1}{2}} & = & \left(\M I+\frac{\nu\D t}{2}\M L_{\V v}\right)\V v^{n}+\left(2\nu\D t\right)^{\frac{1}{2}}\M D_{\V w}\V W_{\V v}^{n}\nonumber \\
\M D\V v^{n+1} & = & 0.\label{eq:incompressible_CN_linear}
\end{eqnarray}
This coupled velocity-pressure \emph{Stokes linear system} can be
solved efficiently even in the presence of non-periodic boundaries
by using a preconditioned Krylov iterative solver, as described in
detail in Ref. \cite{NonProjection_Griffith}. The scheme (\ref{eq:incompressible_CN_P})
reproduces the static covariance of the velocity fluctuations exactly,
$\M C_{\V v}=\M{\Set P}$, for \emph{any} time step. A more intuitive
approach to analysing the scheme (\ref{eq:incompressible_CN_linear})
based on computing the modes of the spatial discretization is described
in Appendix \ref{AppendixModeAnalysis}.

Having illustrated how the implicit midpoint rule (\ref{eq:Crank-Nicolson})
can be applied to the time-dependent Stokes equations, it is simple
to modify the implicit-explicit predictor-corrector schemes described
in Section \ref{sec:IMEX} to account for incompressibility. In Ref.
\cite{LLNS_Staggered} the implicit trapezoidal predictor-corrector
scheme (\ref{eq:impl_trapezoidal}) was used to solve (\ref{eq:fluct_NS_P},\ref{eq:fluct_c_eq}),
but only tested in an essentially linearized context. Here we also
consider the implicit midpoint predictor-corrector scheme (\ref{eq:impl_midpoint}),
for which the predictor stage consists of solving the linear system
for $\tilde{\V v}^{n+\frac{1}{2}}$, $\tilde{\V{\pi}}^{n+\frac{1}{2}}$
and $\tilde{\V c}^{n+\frac{1}{2}}$,
\begin{eqnarray*}
\left(\M I-w_{1}\nu\D t\,\M L_{\V v}\right)\tilde{\V v}^{n+\frac{1}{2}}+\D t\,\M G\tilde{\V{\pi}}^{n+\frac{1}{2}} & = & \frac{\Delta t}{2}\M S_{\V v}\left(\V v^{n}\right)\V v^{n}+\left[\M I+\left(\frac{1}{2}-w_{1}\right)\nu\D t\,\M L_{\V v}\right]\V v^{n}+\left(\nu\D t\right)^{\frac{1}{2}}\M D_{\V w}\left(\V W_{\V v}^{n}\right)_{1}\\
\left(\M I-w_{1}\chi\D t\,\M L_{c}\right)\tilde{\V c}^{n+\frac{1}{2}} & = & \frac{\Delta t}{2}\M S_{c}\left(\V v^{n}\right)\V c^{n}+\left[\M I+\left(\frac{1}{2}-w_{1}\right)\chi\D t\,\M L_{c}\right]\V c^{n}+\left(\epsilon\chi\D t\right)^{\frac{1}{2}}\M D\left(\V W_{c}^{n}\right)_{1}\\
\M D\tilde{\V v}^{n+\frac{1}{2}} & = & 0,
\end{eqnarray*}
and similarly for the corrector stage. The concentration equation
is decoupled from the velocity equation and can be solved using standard
techniques, e.g., multigrid methods. Note that in practice it is better
to rewrite the linear systems in terms of the increments $\tilde{\V v}^{n+\frac{1}{2}}-\V v^{n}$
and $\tilde{\V c}^{n+\frac{1}{2}}-\V c^{n}$. This is because for
weak fluctuations the terms involving the identity matrix may dominate
the right-hand side and compromise the accuracy of the linear solvers,
unless special care is taken in choosing the termination criteria
for the iterative linear solvers.

\section{\label{sec:Numerical-Results}Simulations}

In this section we numerically study the behavior of the implicit-explicit
trapezoidal (\ref{eq:impl_trapezoidal}) and midpoint (\ref{eq:impl_midpoint})
schemes on the fluctuating Burgers and Navier-Stokes equations. We
focus here on the behavior of the equilibrium distribution of the
fluctuating fields for large fluctuations and large time step sizes.
The discrete spectrum of the equilibrium fluctuations is one of the
most important properties of a numerical scheme for long-time simulations.

Our spatial discretizations were constructed to obey a discrete fluctuation-dissipation
balance principle, which means that for sufficiently small time steps
the numerical schemes will produce the correct equilibrium fluctuations
even when the nonlinear terms are important. In the absence of advection,
the schemes (\ref{eq:impl_trapezoidal}) and (\ref{eq:impl_midpoint})
reduce to the implicit midpoint rule, which was designed to produce
the correct equilibrium fluctuations for \emph{any} time step. It
is not \emph{a priori} obvious how increasing the time step size affects
the long-time behavior of schemes in the presence of the nonlinear
advective terms, which are treated explicitly. We study this question
numerically in this section.

In fluctuating hydrodynamics, the magnitude of the equilibrium fluctuations
is controlled by the degree of coarse graining, more specifically,
by the average number of particles $N_{p}$ (microscopic degrees of
freedom) per hydrodynamic cell (macroscopic degree of freedom). In
particular, the law of large numbers suggests that the Gaussian fluctuations
at equilibrium have a variance inversely proportional to $N_{p}\sim\D V$.
In order to model the effect of the degree of coarse graining we can
introduce a parameter $\epsilon\sim N_{p}^{-1}$ that measures the
strength of the fluctuations, with $\epsilon\sim1$ indicating very
strong fluctuations, i.e., minimal coarse graining. While we cannot
expect Markovian SPDE models to be a good approximation to reality
in the absence of coarse-graining, from a numerical analysis perspective
it is important to understand how robust the numerical schemes are
to increased magnitude of the fluctuations. We note that the types
of schemes we use here may have uses in other fields such as turbulence,
where reproducing the correct spectrum of fluctuations is also important.

\subsection{\textmd{\label{sub:NumericsBurgers}}Fluctuating Burgers Equation}

We first turn our attention to the fluctuating Burgers equation. For
simplicity, we take $c=1$ and consider a periodic system with zero
total momentum (no macroscopic advection),
\[
\partial_{t}u+u\,\partial_{x}u=\nu\,\partial_{xx}^{2}u+\left(2\epsilon\nu\right)^{\frac{1}{2}}\,\partial_{x}\mathcal{Z}.
\]
When the spatial discretization (\ref{eq:semidiscrete_Burgers}) is
used, the coarse-grained velocities $u_{j}$ have Gaussian equilibrium
fluctuations with mean zero and covariance
\[
\av{u_{i}u_{j}}=\frac{\epsilon}{\D x}\delta_{ij}.
\]
The equilibrium magnitude of the velocities is therefore $\abs u\approx\sqrt{\epsilon/\D x}$,
which is a measure of the typical magnitude of the advection speed.

\subsubsection{Dimensionless Numbers}

In deterministic fluid dynamics, the dimensionless number that describes
how well advection is resolved by the time step size is the advective
CFL (Courant-Friedrichs-Levy) number
\[
\alpha=\frac{\abs u\D t}{\D x}\approx\D t\epsilon^{\frac{1}{2}}\D x^{-\frac{3}{2}}.
\]
Traditional wisdom says that for schemes that handle advection explicitly
$\D t$ should be chosen such that $\alpha\lesssim1$. The presence
of diffusion, however, stabilizes numerical schemes by adding dissipation
and introduces the viscous CFL number $\beta$ and the cell Reynolds
number $r$,
\[
\beta=\frac{\nu\D t}{\D x^{2}},\quad r=\frac{\alpha}{\beta}=\frac{\abs u\D x}{\nu}\approx\epsilon^{\frac{1}{2}}\nu^{-1}\D x^{\frac{1}{2}}.
\]
Note that the cell Reynolds number measures the relative importance
of the nonlinear term (advection) versus the linear term (diffusion),
and is independent of $\D t$. In the deterministic setting, standard
von Neumann stability analysis for the Euler scheme applied to the
advection-diffusion equation suggests that our centered discretization
of advection is stable if $\alpha^{2}/2\leq\beta\leq1/2$. This suggests
that the dimensionless number
\[
\gamma=\frac{\alpha^{2}}{\beta}=r^{2}\beta=\frac{\abs u^{2}\D t}{\nu}\approx\frac{\epsilon\D t}{\nu\D x}
\]
may be important in controlling the behavior of our spatio-temporal
discretizations.

For weak fluctuations ($\epsilon\ll1$) or strong dissipation ($r\ll1$),
both of which are typically true for realistic fluids at small scales
and sufficient levels of coarse-graining ($N_{p}\gg1$), the accuracy
is controlled by the viscous CFL number $\beta$. In particular, if
$\beta\ll1$ we can be confident that the numerical scheme resolves
the dynamics of the fluctuations accurately. However, running with
small viscous CFL numbers is often impractical. Our semi-implicit
schemes are designed to be stable and also to correctly reproduce
the equilibrium fluctuations even for large time steps, $\beta\gg1$,
at least as long as $\alpha\ll1$. An important question, which is
difficult to analyze with existing analytical tools, is how large
the time step can be before the explicit handling of the nonlinear
advective terms introduces large errors. We study this question here
by examining the equilibrium fluctuations for strong fluctuations,
$\epsilon\gtrsim1$, and large time steps, $\beta\gg1$.

\subsubsection{Static Structure Factors}

In order to study the behavior of the equilibrium fluctuations we
follow the approach used in Ref. \cite{LLNS_S_k}. Instead of studying
the fluctuations in the actual (real space) variables $u_{j}$, we
study the equilibrium discrete Fourier spectrum of the fluctuating
variables, defined as
\[
S_{\kappa}=N\epsilon^{-1}\D x\,\av{\hat{u}_{k}\hat{u}_{k}^{\star}}.
\]
Here the discrete Fourier transform is defined as 
\[
\hat{u}_{k}=N^{-1}\sum_{j=0}^{N}u_{j+1}e^{-ij\D k},
\]
where $0\leq\kappa\leq\left\lfloor N/2\right\rfloor $ is the waveindex
and $\D k=2\pi\kappa/N\leq\pi$ is the dimensionless wavenumber. The
quantity $S_{\kappa}$ is a dimensionless discrete version of what
is called the \emph{static structure factor} $S(k)$ in the physics
literature, where $k=\D k/\D x$ is the physical wavenumber. For fluctuating
hydrodynamics, at thermodynamic equilibrium
\[
S_{\kappa}=1\mbox{ for all }\kappa\neq0,
\]
which is a re-statement of the discrete fluctuation-dissipation balance
principle.

For periodic systems, due to translational invariance, the quantity
$S_{\kappa}$ contains the same statistics about the equilibrium fluctuations
as the $N\times N$ covariance matrix $C_{j,j^{\prime}}=\av{u_{j}u_{j^{\prime}}}$.
The advantage of using the Fourier description is that it illustrates
the behavior at different physical length scales. It is expected that
any numerical scheme will produce some artifacts at the largest wavenumbers
because of the strong corrections due to the discretization; however,
small wavenumbers, $\D k\ll1$, ought to have much smaller errors
because they evolve over time scales and length scales much larger
than the discretization step sizes $\D x$ and $\D t$. A scheme or
choice of time step size that produces a \emph{discrete structure}
\emph{factor} $S_{\kappa}$ much different from unity at small wavenumbers
must be rejected as unphysical. It is important to emphasize, however,
that getting a good equilibrium \emph{static} spectrum for the fluctuations
is not a guarantee that a scheme accurately models the \emph{dynamics}
of the fluctuations.

\subsubsection{Numerical Results}

\begin{figure}
\begin{centering}
\includegraphics[width=0.49\textwidth]{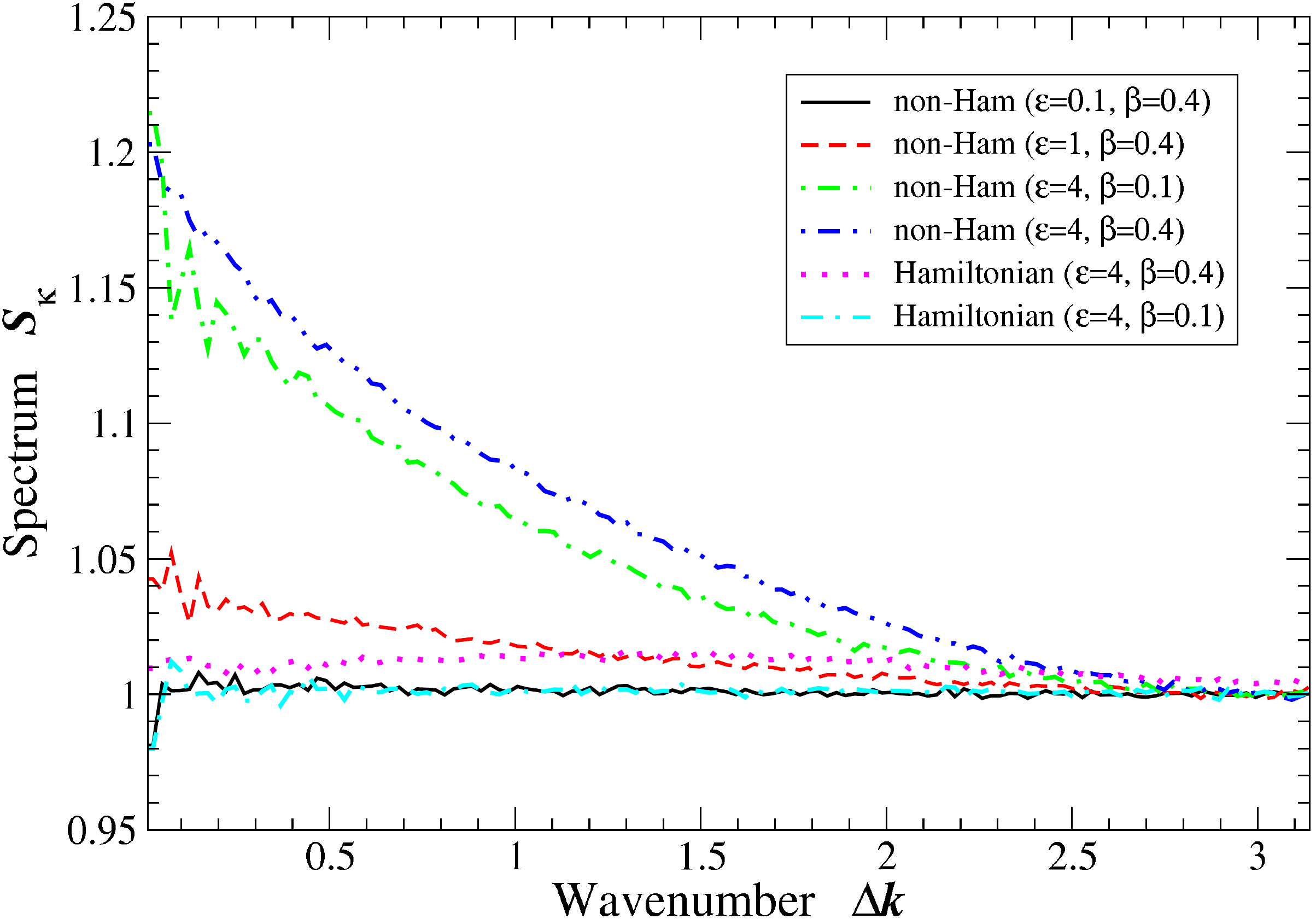}\includegraphics[width=0.49\textwidth]{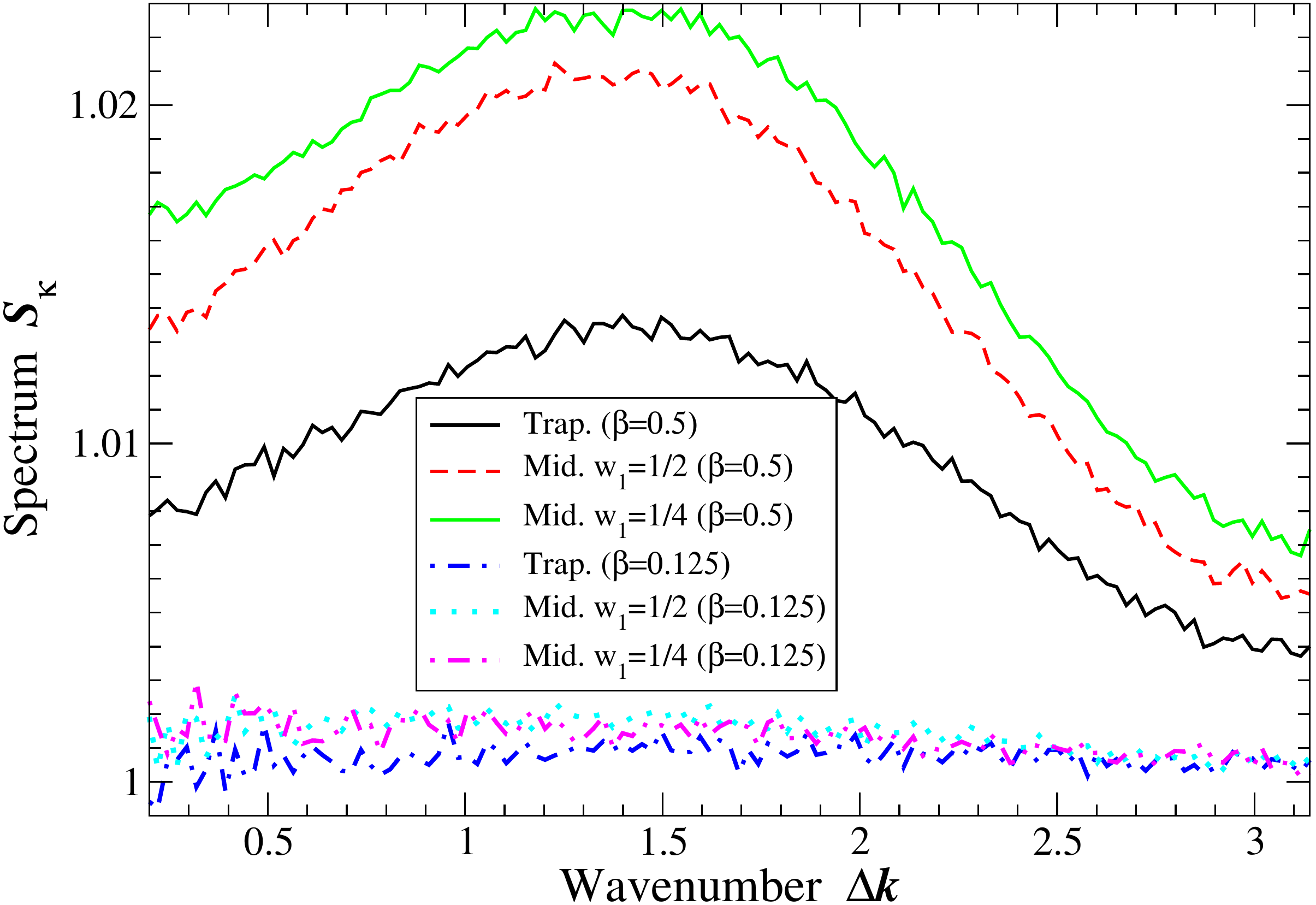}
\par\end{centering}

\caption{\label{fig:Burgers_large_fluct_S_k} Static structure factor $S_{\kappa}$
for a periodic system of 256 cells at thermodynamic equilibrium. (\emph{Left
panel}) Midpoint scheme (\ref{eq:impl_midpoint}) with $w_{1}=1/2$
and non-Hamiltonian advection (\ref{eq:advection_Burgers_nonH}) for
time step size $\beta=0.4$ for weak fluctuations $\epsilon=0.1$
($\alpha\approx0.13$), medium fluctuations $\epsilon=1$ ($\alpha=0.4$)
and strong fluctuations $\epsilon=4$ ($\alpha=0.8$). For $\epsilon=4$,
reducing the time step to $\beta=0.1$ ($\alpha=0.2$) does not significantly
improve the accuracy and even if $\D t\rightarrow0$ the wrong spectrum
of fluctuations is obtained. Switching to the Hamiltonian discretization
(\ref{eq:advection_Burgers}) significantly lowers the errors. (\emph{Right
panel}) Comparison between the trapezoidal scheme (\ref{eq:impl_trapezoidal})
and the midpoint scheme (\ref{eq:impl_midpoint}) with $w_{1}=1/2$
and with $w_{1}=1/4$, for $\epsilon=4$ and large time step size
$\beta=0.5$ ($\alpha=1$) and small time step $\beta=0.125$ ($\alpha=0.25$).
Advection is discretized in a Hamiltonian manner.}
\end{figure}

In Fig. \ref{fig:Burgers_large_fluct_S_k} we show numerical results
for the equilibrium structure factor $S_{\kappa}$ for a periodic
system with $\nu=1$ and $\D x=1$ and zero total momentum, $\av{u_{j}}=0$.
To illustrate the importance of using a Hamiltonian discretization
of advection in the nonlinear setting, we consider a scheme where
the advective term $uu_{x}$ is handled using the conservative but
non-Hamiltonian discretization
\begin{equation}
\left(\M S\V u\right)_{j}=-\frac{c}{2}\left(\frac{u_{j+\frac{1}{2}}^{2}-u_{j-\frac{1}{2}}^{2}}{\D x}\right)=-\frac{c}{2}\left(\frac{u_{j+1}^{2}-u_{j-1}^{2}}{2\D x}\right),\mbox{ where }u_{j+\frac{1}{2}}^{2}=\frac{u_{j}^{2}+u_{j+1}^{2}}{2},\label{eq:advection_Burgers_nonH}
\end{equation}
instead of the conservative Hamiltonian discretization (\ref{eq:advection_Burgers}).
We recall that the correct answer is $S_{\kappa}=1$ for all wavenumbers.
The results in the left panel of Fig. \ref{fig:Burgers_large_fluct_S_k}
illustrate that for weak fluctuations (i.e., nearly linear equations),
the correct spectrum is obtained. However, for strong fluctuations
(i.e., nonlinear equations), $\epsilon=4$, the non-Hamiltonian scheme
produces the wrong static spectrum of fluctuations at small wavenumbers,
and reducing $\D t$ does not help.

On the other hand, the Hamiltonian discretization gives small errors
in the spectrum even for the larger time step size. In the right panel
of Fig. \ref{fig:Burgers_large_fluct_S_k} we zoom in to show the
magnitude and form of the errors in the structure factors for the
implicit-explicit trapezoidal scheme (\ref{eq:impl_trapezoidal})
and for the midpoint scheme (\ref{eq:impl_midpoint}) with $w_{1}=1/2$
and with $w_{1}=1/4$. We see that all three schemes show similarly
small errors in the spectrum. Reducing $\D t$ by a factor of four
makes the error statistically insignificant. We have verified that
the errors are of second order in the time step size $\D t$ for all
three schemes.

\begin{figure}
\begin{centering}
\includegraphics[width=0.49\textwidth]{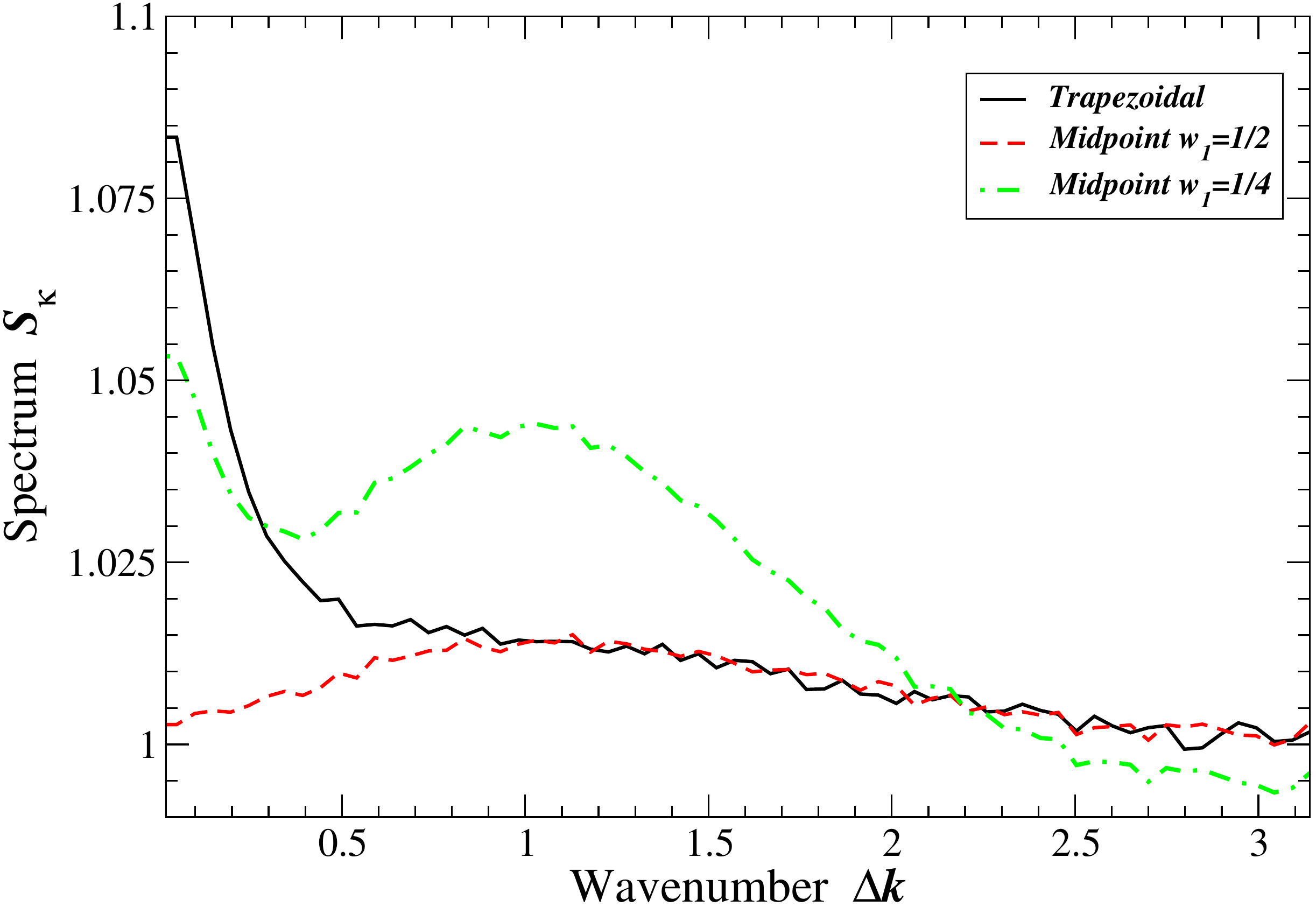}\includegraphics[width=0.49\textwidth]{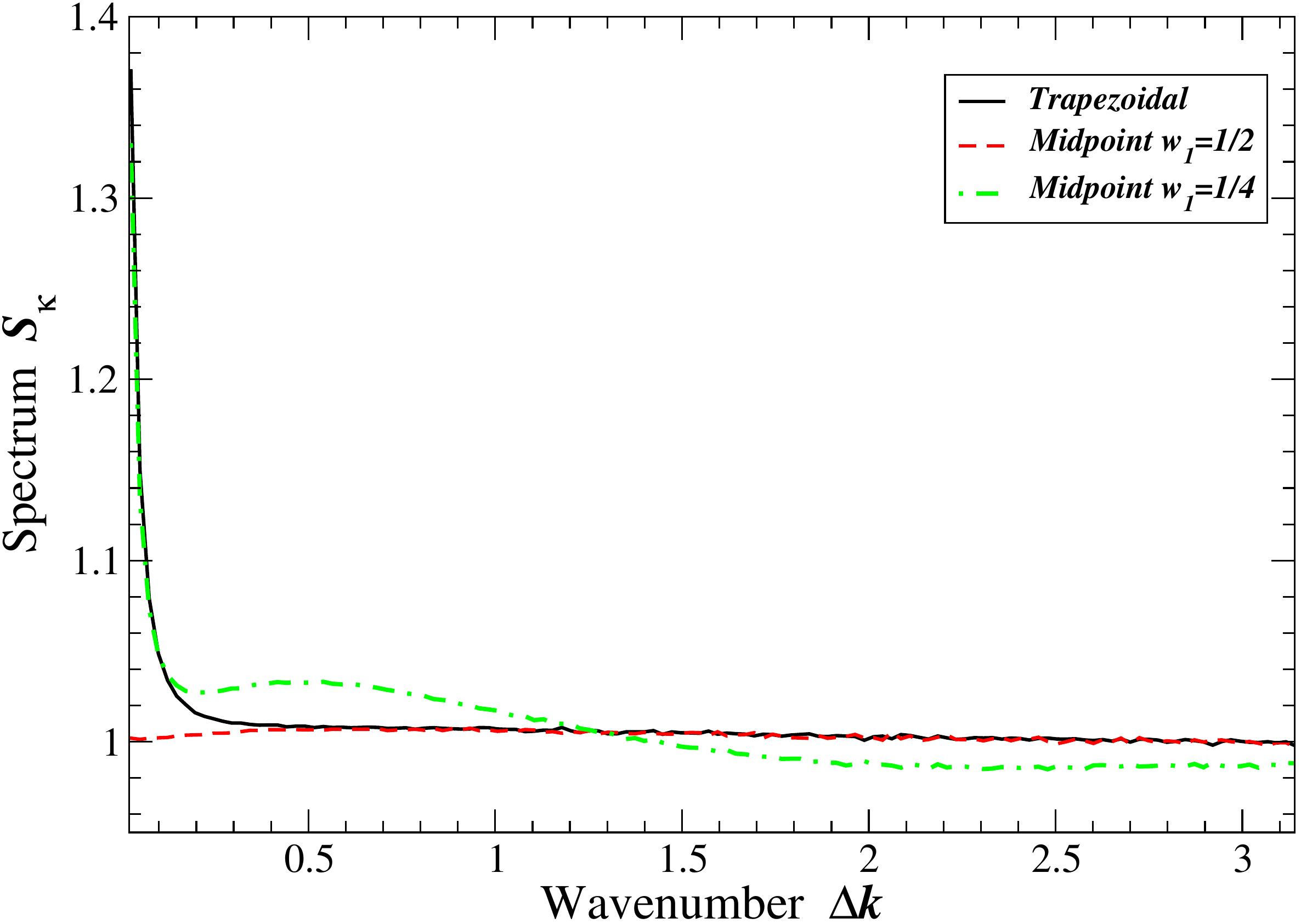}
\par\end{centering}

\caption{\label{fig:Burgers_weak_fluct_S_k}Static structure factor $S_{\kappa}$
at thermodynamic equilibrium in the case of weaker fluctuations and
large time step sizes, for the trapezoidal scheme (\ref{eq:impl_trapezoidal})
and for the midpoint scheme (\ref{eq:impl_midpoint}) with $w_{1}=1/2$
and with $w_{1}=1/4$. (\emph{Left panel}) Moderate fluctuations,
$\epsilon=0.1$, for time step size $\beta=10$, $\alpha\approx3.2$,
$\gamma=1$. (\emph{Right panel}) Weak fluctuations, $\epsilon=0.01$
and time time step size, $\beta=100$, $\alpha=10$, $\gamma=1$.}
\end{figure}

In Fig. \ref{fig:Burgers_weak_fluct_S_k} we show numerical results
for the static structure factor in the case of weaker fluctuations,
$\epsilon=0.1$ and $\epsilon=0.01$, but large time step size. This
is a typical scenario for fluctuating hydrodynamics in practice, since
for reasonable degree of coarse graining the fluctuations would be
small and the behavior of the equations would be close to that of
the linearized equations. In the absence of the advective nonlinearity
our schemes are stable for arbitrary viscous CFL number $\beta$.
Under equilibrium conditions the semi-implicit predictor-corrector
schemes we consider are observed to be stable up to rather large $\D t$
as measured in the advective CFL $\alpha$, and even the dimensionless
number $\gamma=\alpha^{2}/\beta$. However, for sufficiently large
$\D t$ the nonlinearities are expected to play some role, and one
cannot expect to be able to increase the time step size up to the
stability limit and still maintain reasonable accuracy. The results
in Fig. \ref{fig:Burgers_weak_fluct_S_k} show that for large $\D t$
there appear significant artifacts in the static structure factor
for the trapezoidal scheme (\ref{eq:impl_trapezoidal}) and for the
midpoint scheme (\ref{eq:impl_midpoint}) with $w_{1}=1/4$. While
the magnitude of the errors is small, the problematic observation
is that the errors have a peak at the smallest wavenumbers, where
we expect schemes to most closely mimic the continuum equations.

The above observations lead us to select the midpoint scheme (\ref{eq:impl_midpoint})
with $w_{1}=1/2$ as the most robust temporal integrator for the fluctuating
Burgers equation. At the same time, we should recognize that the best
choice of scheme will depend on the quantity of interest and the particular
problem under consideration. All schemes are observed to produce equilibrium
fluctuations that are rather robust under the presence of strong non-linearities
if a Hamiltonian advection of discretization is employed. In the next
section we confirm that this conclusion also holds for the fluctuating
Navier-Stokes equation.

\subsection{Fluctuating Navier-Stokes Equation with Passive Tracer}

We now turn our attention to the fluctuating Navier-Stokes equations
with a passively-advected scalar, Eqs. (\ref{eq:fluct_NS_P},\ref{eq:fluct_c_eq}).
Here we set $\rho^{-1}\, k_{B}T=\epsilon$ so that $\epsilon$ controls
the magnitude of both the velocity and the concentration fluctuations.

The implementation of our spatio-temporal discretizations and their
accuracy in the linearized setting (weak fluctuations) is discussed
in more detail in Ref. \cite{LLNS_Staggered}. Our implementation
is integrated into the the IBAMR software framework \cite{IBAMR},
an open-source library for developing fluid-structure interaction
models that use the immersed boundary method. Based on our experience
with the fluctuating Burgers equation, we focus our attention on the
implicit-explicit trapezoidal scheme (\ref{eq:impl_trapezoidal})
and the midpoint scheme (\ref{eq:impl_midpoint}) with $w_{1}=1/2$.
For simplicity, here we focus on two spatial dimensions and periodic
boundary conditions, but we wish to emphasize that our formulation,
numerical schemes, and implementation apply to three spatial dimensions
and no-slip or free-slip boundaries as well.

For simplicity, in our numerical tests we set $\D x=\D y=1$. The
diffusion coefficients for momentum and concentration are set to $\nu=1$
and $\chi=0.25$, and a grid of size $64\times64$ is employed. The
same dimensionless numbers as for the Burgers equation apply, with
the difference that there is a separate diffusive CFL for the concentration,
$\beta_{c}=\chi\beta/\nu$, and therefore the cell Peclet number $r_{c}=\abs u\D x/\chi$
is four times larger than the cell Reynolds number $r$.

\subsubsection{Static Structure Factors}

The equilibrium fluctuations in velocity and concentration are characterized
by the static structure factors, which are the equilibrium average
of the discrete Fourier spectrum of the fluctuating velocities and
concentrations. Concentration fluctuations are characterized via
\[
S_{\V{\kappa}}^{(c)}=N_{c}\epsilon^{-1}\D V\,\left\langle \hat{c}_{\V{\kappa}}\hat{c}_{\V{\kappa}}^{\star}\right\rangle ,
\]
where $\D V=\D x\,\D y$ is the volume of the hydrodynamic cells,
$N_{c}$ is the number of hydrodynamic cells, and $\V{\kappa}=\left(\kappa_{x},\kappa_{y}\right)$
is the waveindex. For the velocity fluctuations, we calculate the
spectrum of the fluctuations of a variable related to vorticity \cite{LLNS_Staggered},
\[
S_{\V{\kappa}}^{(\Omega)}=N_{c}\epsilon^{-1}\D V\,\left\langle \hat{\Omega}_{\V{\kappa}}\hat{\Omega}_{\V{\kappa}}^{\star}\right\rangle ,
\]
where $\hat{\Omega}_{\V{\kappa}}$ is obtained from the discrete Fourier
spectrum of the velocity components as 
\[
\hat{\Omega}_{\V{\kappa}}=k^{-1}\left(k_{x}\hat{v}_{y}-k_{y}\hat{v}_{x}\right),
\]
and $k=\sqrt{k_{x}^{2}+k_{y}^{2}}$ is the wavenumber. Note that $S_{\V{\kappa}}^{(\Omega)}$
fully characterizes the covariance of the velocity fluctuations since
our scheme ensures the velocity is discretely divergence free at all
times and $k^{-1}\left(-k_{y},\, k_{x}\right)$ spans the subspace
of divergence-free velocities in Fourier space. For staggered variables
the shift between the corresponding grids should be taken into account
as a phase shift in Fourier space, for example, $\exp\left(k_{x}\D x/2\right)$
for $v_{x}$. Additionally, the wavenumber $\V k=\left(k_{x},k_{y}\right)$
should be replaced by the effective wavenumber $\tilde{\V k}$ that
takes into account the centered discretization of the projection operator,
for example,
\begin{equation}
\tilde{k}_{x}=\frac{\exp\left(ik_{x}\D x/2\right)-\exp\left(-ik_{x}\D x/2\right)}{i\D x}=k_{x}\frac{\sin\left(k_{x}\D x/2\right)}{\left(k_{x}\D x/2\right)}.\label{eq:k_x_tilde}
\end{equation}
One can additionally define and measure the cross-correlation between
concentration and velocity fluctuations via the cross-correlation
static structure factor
\[
S_{\V{\kappa}}^{(c,\Omega)}=N_{c}\epsilon^{-1}\D V\,\left\langle \hat{c}_{\V{\kappa}}\hat{\Omega}_{\V{\kappa}}^{\star}\right\rangle ,
\]
which is in general a complex number.

Discrete fluctuation-dissipation balance in the coupled velocity-concentration
equations requires that $S_{\V{\kappa}}^{(c)}=S_{\V{\kappa}}^{(\Omega)}=1$
and $S_{\V{\kappa}}^{(c,\V v)}=0$ for all nonzero wavenumbers. Deviations
from these values indicate a violation of discrete fluctuation-dissipation
balance and can be used to numerically assess the behavior of the
schemes in the nonlinear setting, as we do next.

\subsubsection{Numerical Results}

\begin{figure*}
\begin{centering}
\includegraphics[width=0.45\textwidth]{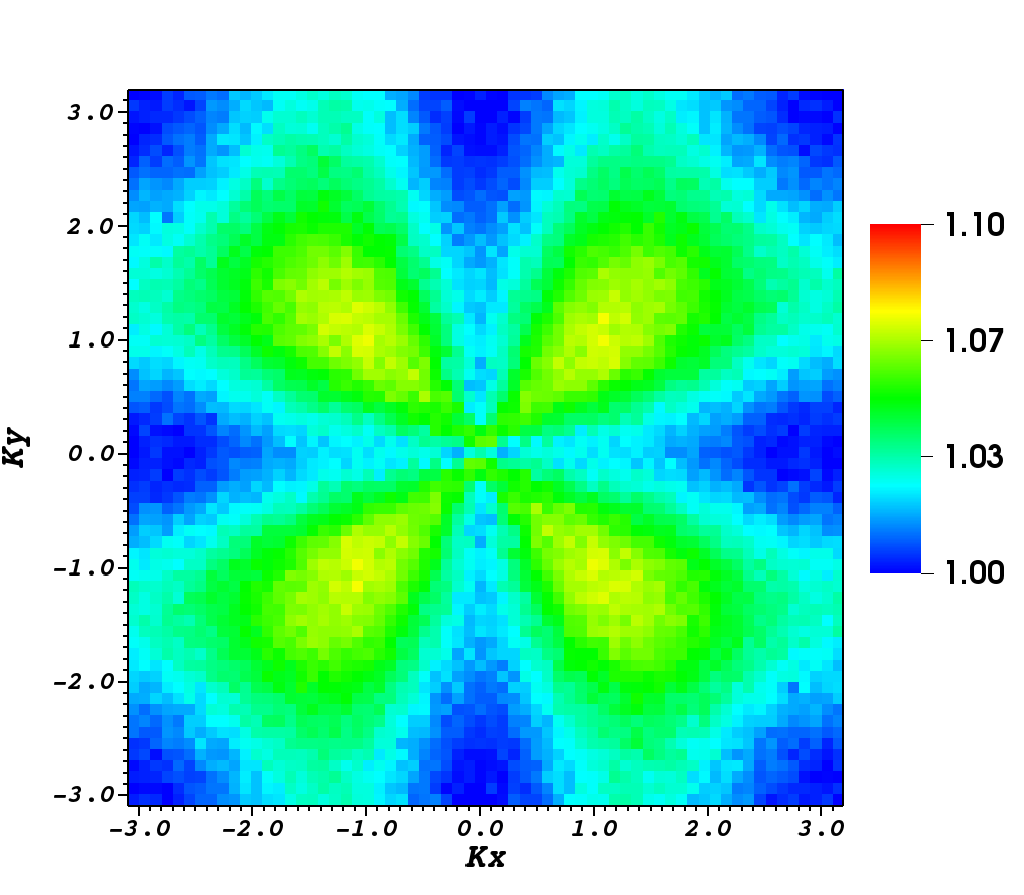}\includegraphics[width=0.45\textwidth]{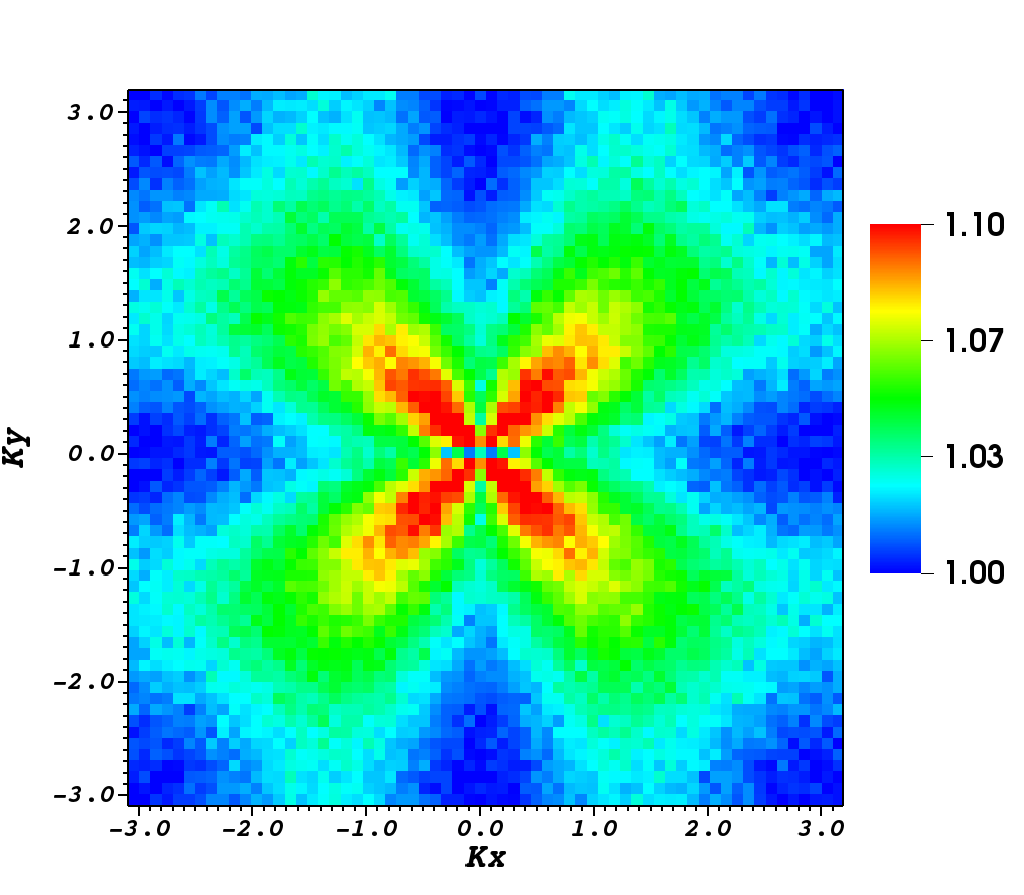}
\par\end{centering}

\begin{centering}
\includegraphics[width=0.45\textwidth]{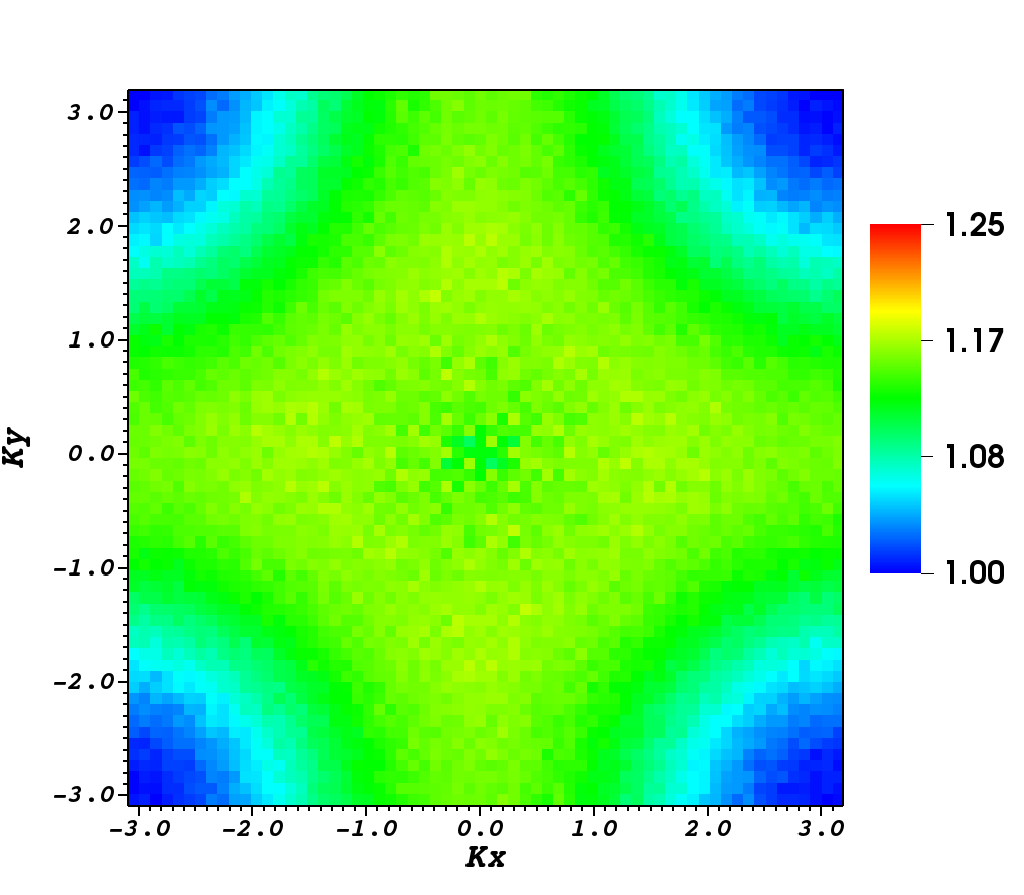}\includegraphics[width=0.45\textwidth]{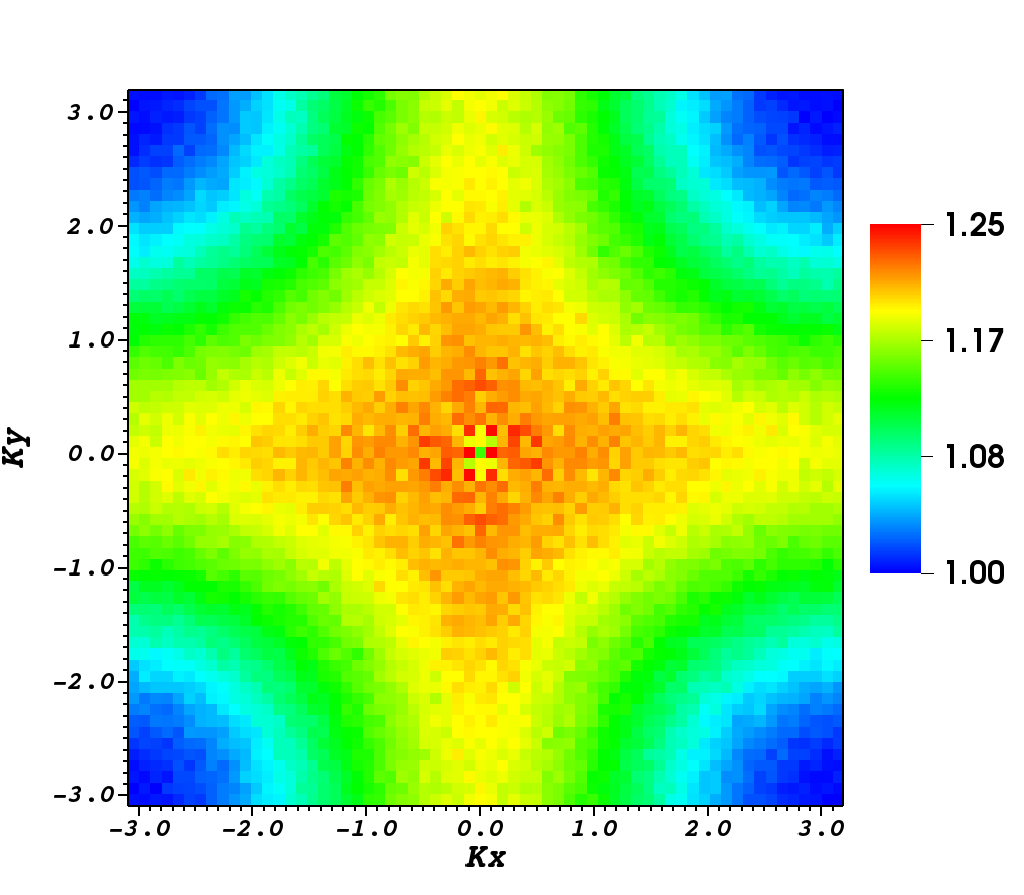}
\par\end{centering}

\caption{\label{fig:S_c_v_large_fluct}Discrete structure factors $S_{\V{\kappa}}^{(\Omega)}$
(top) and $S_{\V{\kappa}}^{(c)}$ (bottom) for the midpoint scheme
(\ref{eq:impl_midpoint}) with $w_{1}=1/2$ on the left and for the
trapezoidal scheme (\ref{eq:impl_trapezoidal}) on the right, for
the case of large fluctuations $\epsilon=4$. (\emph{Top row}) Velocity
spectra for a time step size $\beta=0.5$, $\alpha=1$. (\emph{Bottom
row}) concentration spectra for $\beta_{c}=\chi\beta/\nu=0.0625$
and $\alpha=0.5$.}
\end{figure*}

In Fig. \ref{fig:S_c_v_large_fluct} we show numerical results for
the spectrum of the fluctuations in the solenoidal modes of velocity
and concentration for strong fluctuations, $\epsilon=4$. We see that,
just as for the fluctuating Burgers equation, both the trapezoidal
and the midpoint scheme show artifacts in the spectra, especially
for concentration and for the trapezoidal scheme. The cross-correlation
$S_{\V{\kappa}}^{(c,\Omega)}$ is found to be small and difficult
to measure due to large statistical errors.

\begin{figure}
\begin{centering}
\includegraphics[width=0.49\textwidth]{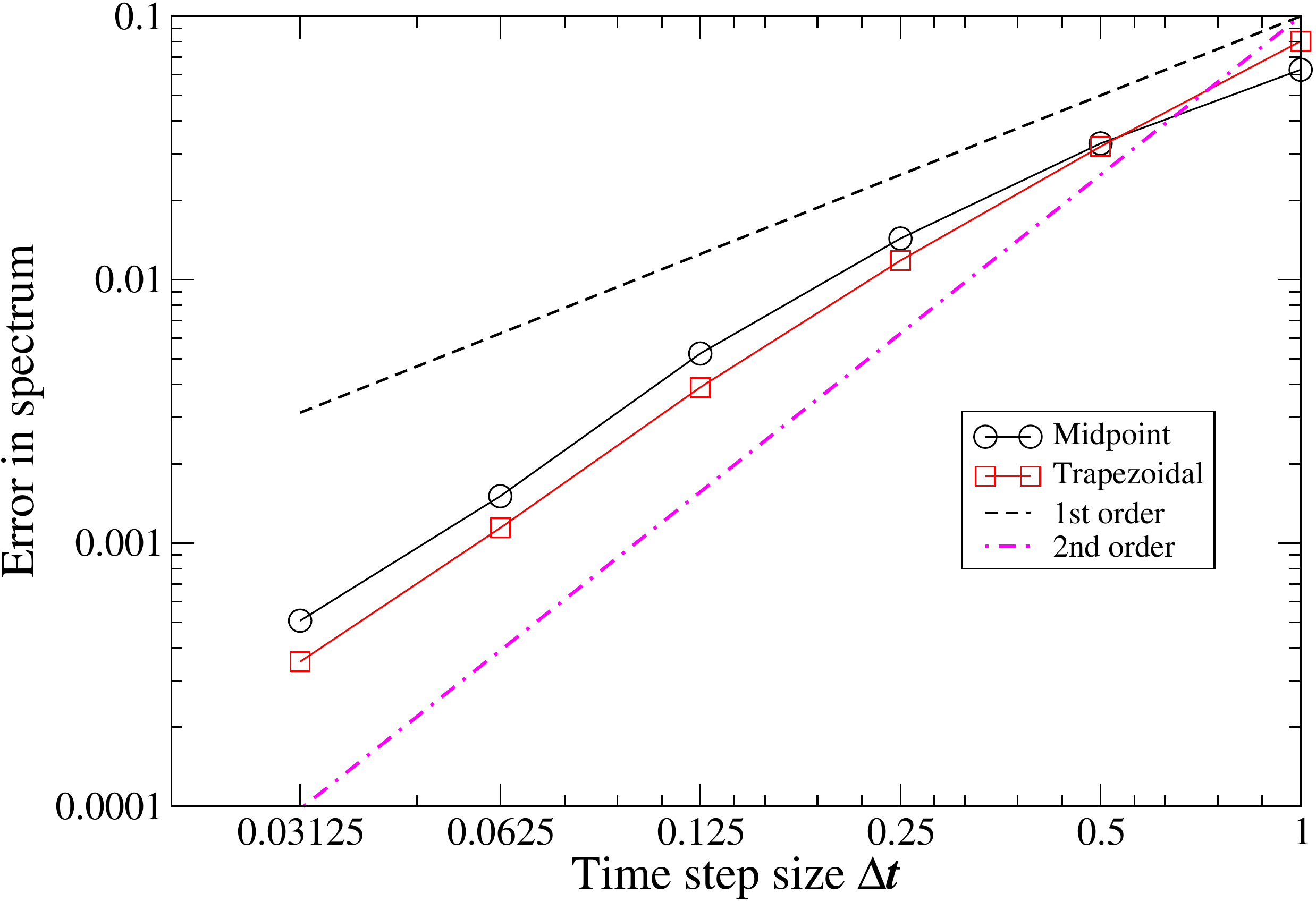}\includegraphics[width=0.49\textwidth]{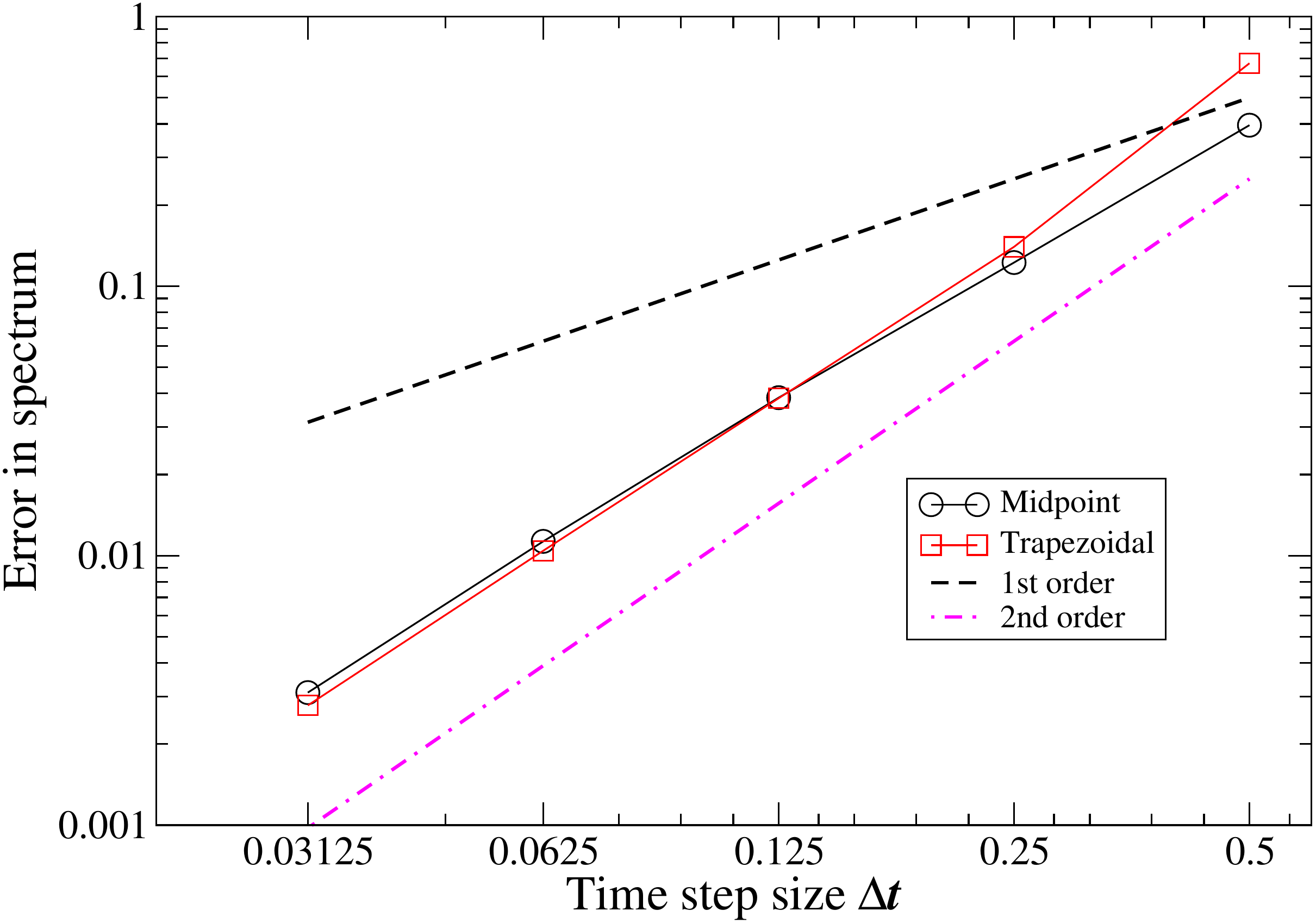}
\par\end{centering}

\caption{\label{fig:S_k_error_NS}Average error in the static spectra for several
time step sizes for the trapezoidal scheme (\ref{eq:impl_trapezoidal})
and for the midpoint scheme (\ref{eq:impl_midpoint}) with $w_{1}=1/2$,
for the case of large fluctuations $\epsilon=4$. First- and second-order
error trends are indicated, showing the second-order asymptotic accuracy.
Error bars are comparable to the symbol size and not shown. (Left
panel) $\av{S_{\V{\kappa}}^{(\Omega)}-1}$. (Right panel) $\av{S_{\V{\kappa}}^{(c)}-1}$.}
\end{figure}

We have verified that as the time step is reduced, both schemes give
the correct spectrum even for strong fluctuations (i.e., strong nonlinearities).
In Fig. \ref{fig:S_k_error_NS} we show the average error in the equilibrium
spectrum $\av{\abs{S_{\V{\kappa}}-1}}$ for vorticity and concentration.
The second-order weak accuracy of the error is clearly seen in Fig.
\ref{fig:Burgers_weak_fluct_S_k} for both velocity and concentration
and for both the midpoint and the trapezoidal scheme.

\begin{figure*}
\begin{centering}
\includegraphics[width=0.45\textwidth]{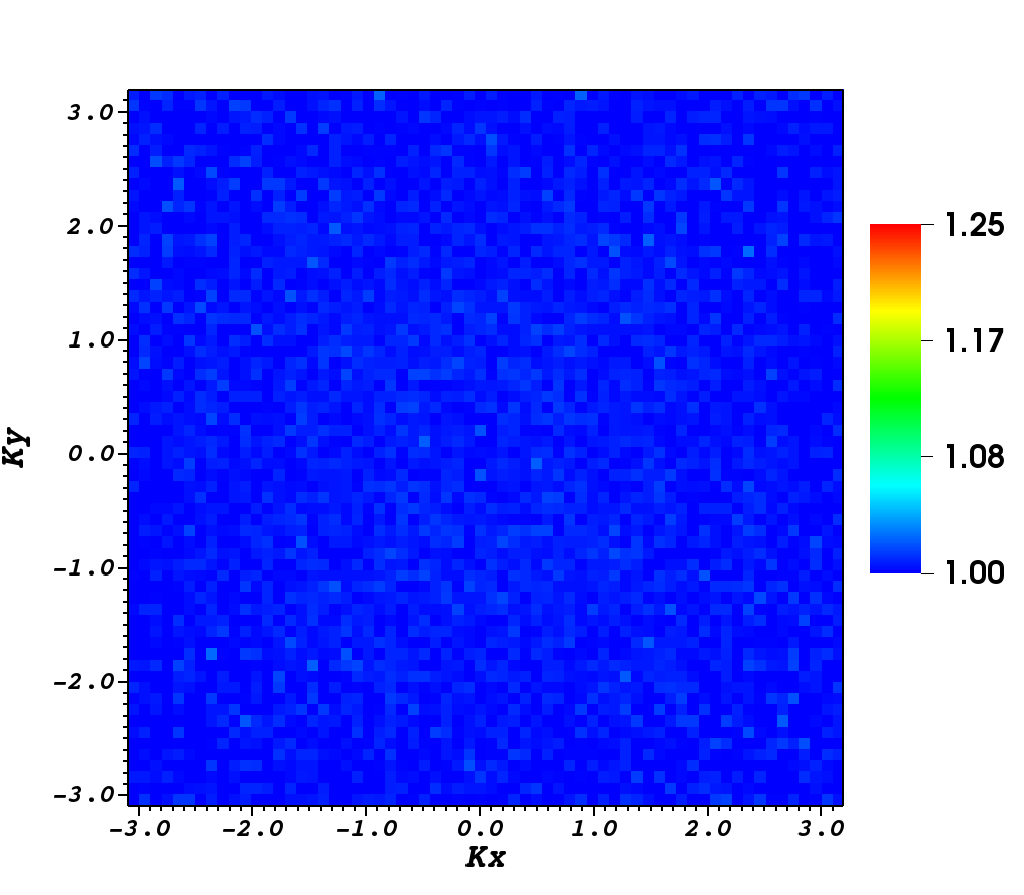}\includegraphics[width=0.45\textwidth]{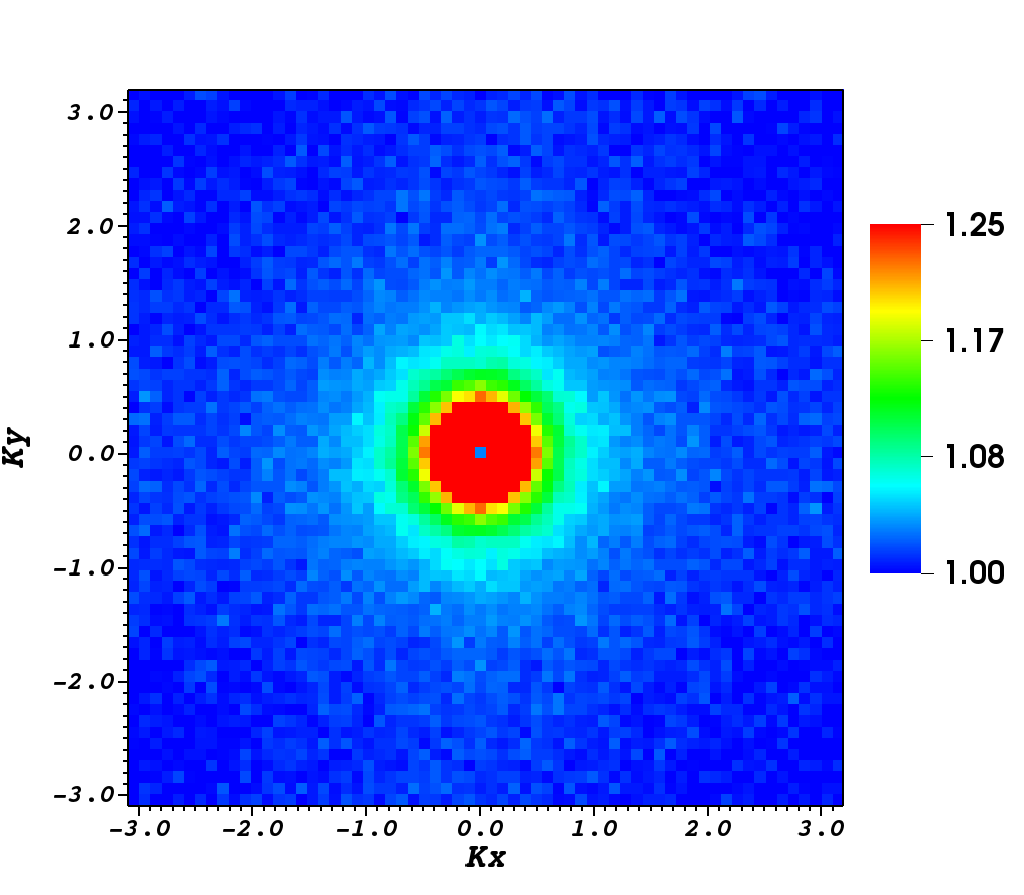}
\par\end{centering}

\caption{\label{fig:S_c_weak_fluct}Discrete structure factor for concentration
$S_{\V{\kappa}}^{(c)}$ for the midpoint scheme (\ref{eq:impl_midpoint})
with $w_{1}=1/2$ on the left and for the trapezoidal scheme (\ref{eq:impl_trapezoidal})
on the right, for the case of weak fluctuations, $\epsilon=0.01$,
and large time step size $\beta=50$, $\beta_{c}=12.5$, $\alpha=5$.}
\end{figure*}

In Fig. \ref{fig:S_c_weak_fluct} we show the spectrum of concentration
fluctuations $S_{\V{\kappa}}^{(c)}$ for the case of weak fluctuations,
$\epsilon=0.01$, and large time step size, viscous CFL $\beta=50$
and diffusive CFL $\beta_{c}=12.5$, and advective CFL $\alpha=5$.
We see a large error for small wavenumbers for the trapezoidal scheme.
A similar but weaker artifact is seen for the velocity spectrum $S_{\V{\kappa}}^{(\Omega)}$
as well, indicating that the trapezoidal scheme violates fluctuation-dissipation
balance at small wavenumbers for large $\D t$. The midpoint scheme
is seen to be much more accurate for both $S_{\V{\kappa}}^{(c)}$
and $S_{\V{\kappa}}^{(\Omega)}$. These investigations confirm that
the midpoint scheme (\ref{eq:impl_midpoint}) with $w_{1}=1/2$ is
the more robust temporal integrator for fluctuating hydrodynamics.

\section{\label{sec:Conclusions}Conclusions}

Discrete and continuum Langevin models are often used as coarse-grained
models for the behavior of materials at mesoscopic scales. These models
include the effects of thermal fluctuations via white-noise stochastic
forcing terms chosen in a way that ensures fluctuation-dissipation
balance. This means that at thermodynamic equilibrium the dynamics
is time reversible (i.e., in detailed balance) with respect to the
Gibbs-Boltzmann distribution. For continuum models this is usually
only true formally and a more precise interpretation of the equations
requires introducing a spatial discretization and truncating the continuum
models at a scale well-separated from the molecular scale. Here we
focused on fluctuating hydrodynamics, specifically, we considered
numerical methods for solving the fluctuating Burgers equation in
one dimension and the fluctuating Navier-Stokes equations in two dimensions.
In these equations, fluctuation-dissipation balance is obtained from
the balance of the dissipative (self-adjoint) diffusive terms and
the stochastic forcing. The advection terms are non-dissipative (skew-adjoint)
and do not affect the equilibrium distribution.

Spatio-temporal discretizations do not necessarily preserve the properties
of the continuum equations, notably, they may not obey a discrete
fluctuation-dissipation principle. This means that their equilibrium
distribution is not a discrete form of the Gibbs-Boltzmann distribution.
This sort of unphysical behavior can be avoided by carefully constructing
the spatial discretization to obey a discrete-fluctuation principle
with respect to a target discrete Gibbs-Boltzmann distribution. In
this way, a coarse-grained semi-discrete model is obtained that obeys
the principles of statistical mechanics, namely, obeys detailed balance
at thermodynamic equilibrium. We showed explicitly how one can construct
such spatial discretizations by first understanding the properties
of the continuum model (even if only at a formal level) and then maintaining
those properties in the spatial discretization. For fluctuating hydrodynamics,
this means that certain relations between the discrete divergence,
gradient and Laplacian operators must be preserved, such as the fact
that the divergence and gradient operators are negative adjoints of
each other. We showed how to construct spatial discretizations that
obey a discrete fluctuation-dissipation principle for both the fluctuating
Burgers and the fluctuating Navier-Stokes with the addition of a passively-advected
scalar. These results were mostly a summary of discretizations previously
constructed in somewhat disjoint bodies of literature, and are a stochastic
equivalent of a method-of-lines approach for deterministic fluid dynamics
\cite{FD_Nematics_Adhikari}.

The novel challenge that we tackled here is the construction of temporal
integrators that are weakly second-order accurate for additive noise,
and preserve the fluctuation-dissipation balance reasonably accurately.
It is possible to construct temporal integrators that strictly preserve
the Gibbs-Boltzmann distribution at equilibrium by combining a Metropolis-Hastings
rejection procedure with a more classical temporal integrator. However,
effective use of such Metropolization \cite{MetropolizedSDEs} requires
first constructing a classical weak integrator that produces a good
approximation to the equilibrium distribution for time step sizes
that are only limited by the deterministic dynamics. For separable
Langevin equations, it is well-known how to construct symplectic integrators
that preserve dynamical invariants, including the equilibrium distribution,
very robustly \cite{Hamiltonian_Leimkuhler}. One of the simplest
second-order methods of this type suitable for fluctuating hydrodynamics
is the implicit midpoint rule. However, an implicit discretization
of the nonlinear advective terms is arguably impractical in three
dimensions. Many existing deterministic schemes in fluid dynamics
use a mixed implicit-explicit approach in which the diffusive terms
are handled implicitly and the advective terms are handled explicitly.
We showed how to construct predictor-corrector second-order Runge-Kutta
schemes based on the implicit midpoint rule (Crank-Nicolson method)
for the diffusive terms, and either an explicit midpoint or an explicit
trapezoidal rule for the advective terms. These schemes have the remarkable
property that in the linearized setting (weak fluctuations) they exactly
preserve the Gaussian approximation to the Gibbs-Boltzmann distribution.
We obtained the conditions for second-order weak accuracy of a rather
general class of two-stage Runge-Kutta methods, and proposed several
specific schemes. Through numerical investigations, we found the scheme
(\ref{eq:impl_midpoint}), which combines the implicit and explicit
midpoint rules, was very robust in reproducing the correct spectrum
of fluctuations, even for strong fluctuations (strong nonlinearities)
and for large time step sizes.

A remaining challenge is how to control not just the static distribution
of the equilibrium fluctuations but also to control the accuracy of
the dynamics of the fluctuating fields. This is particularly challenging
for the case in which there are several physical processes with large
separation of time scales, such as, for example, the diffusion of
momentum and of mass. Notably, for many realistic fluids the Schmidt
number $S_{c}=\nu/\chi$ is on the order of $10^{2}-10^{3}$, making
the sort of integrators we discussed impractical because the time
step would be severely limited by the viscosity $\nu$ and not by
the diffusion coefficient $\chi$. Temporal integrators for this sort
of multiscale, or more appropriately, manyscale system of SPDEs, will
be the subject of future investigations. Future work should also explore
Metropolization \cite{MetropolizedSDEs} strategies for predictor-corrector
schemes for fluctuating hydrodynamics, particularly for the case of
multiplicative noise.

We focused here on constructing temporal integrators with weak accuracy
of order two for additive equations. This choice is guided in part
by physical intuition that has yet to be justified more rigorously.
Specifically, fluctuating hydrodynamics is typically applied at length
and time scales where fluctuations are weak and the dynamics is in
a nearly linearized regime. Notably, in many cases of interest the
noise is essentially additive or can be well-approximated as additive.
Weak accuracy is emphasized because it is more relevant in the types
of Monte Carlo applications we are interested in. While the classical
definition of strong order of accuracy is likely too strong for Monte
Carlo simulation, some notion of pathwise convergence is important
in reproducing the statistics of typical paths, such as, for example,
rare transition statistics. Recently, error metrics other than weak
and strong error have also been considered. Notably, Ref. \cite{TotalVariation_SDEs}
defines the numerical error as the difference between the joint PDF
of the computed values at all time steps and the joint PDF of the
exact process at the same times, and constructs Runge-Kutta integrators
accurate in that metric. Such approaches may be fruitfully applied
in fluctuating hydrodynamics in the future, especially in the context
of modeling rare events.
\begin{acknowledgments}
We would like to thank John Bell, Alejandro Garcia, and Jonathan Goodman
for numerous enlightening discussions. A. Donev was supported in part
by the National Science Foundation (NSF) under grant DMS-1115341 and
the Office of Science of the U.S. Department of Energy (DOE) Early
Career award number DE-SC0008271. B. Griffith and S. Delong acknowledge
research support from the NSF under award OCI 1047734. E. Vanden-Eijnden
and S. Delong were supported by the DOE office of Advanced Scientific
Computing Research under grant DE-FG02-88ER25053. Additional support
for E. Vanden-Eijnden was provided by the NSF under grant DMS07-08140
and by the Office of Naval Research under grant N00014-11-1-0345.
\end{acknowledgments}
\begin{appendix}

\section{\label{Appendix-Accuracy}Weak Semi-Implicit Predictor-Corrector
Schemes}

We consider implicit-explicit stochastic Runge-Kutta schemes to solve
the system of Ito SODEs 
\[
d\V x=\boldsymbol{L}(\boldsymbol{x})\boldsymbol{x}\, dt+\boldsymbol{g}(\boldsymbol{x})\, dt+\alpha\,\partial_{\V x}\cdot\left[\boldsymbol{K}(\boldsymbol{x})\boldsymbol{K}^{\star}(\boldsymbol{x})\right]dt+\boldsymbol{K}(\boldsymbol{x})\, d\boldsymbol{\mathcal{B}}(t),
\]
where $\boldsymbol{L}(\boldsymbol{x})$ is a linear operator that
represents the implicitly-treated part of the dynamics. Here $\boldsymbol{\mathcal{B}}$
is a collection of independent Brownian motions (Wiener processes),
with $\V{\mathcal{W}}\equiv d\boldsymbol{\mathcal{B}}/dt$ denoting
a collection of white noise processes. For increased generality we
allow for multiplicative noise and include a divergence term proportional
to some constant $\alpha$, as may be required to ensure fluctuation-dissipation
balance in the case of multiplicative noise. For example, $\alpha=1/2$
gives the spurious or thermal drift term in the generic Langevin equation
(\ref{eq:x_t_general}) if the non-dissipative part of the dynamics
is Hamiltonian. We develop a method that is weakly second order accurate
for additive noise and first order accurate for multiplicative noise,
requires solving only \emph{linear} systems involving the matrix $\M L^{n}=\boldsymbol{L}(\boldsymbol{x}^{n})$,
and does not require evaluating the divergence of the mobility matrix
$\M M=\M K\M K^{\star}$.

We employ a two-stage (predictor-corrector) Runge-Kutta method, where
at time step $n$ the predictor estimates a first-order accurate solution
at an intermediate time $n\D t+w_{2}\D t$, 
\begin{equation}
\tilde{\boldsymbol{x}}=\boldsymbol{x}^{n}+(w_{2}-w_{1})\Delta t\boldsymbol{L}^{n}\boldsymbol{x}^{n}+w_{1}\Delta t\boldsymbol{L}^{n}\tilde{\boldsymbol{x}}+w_{2}\Delta t\boldsymbol{g}^{n}+\left(w_{2}\Delta t\right)^{\frac{1}{2}}\boldsymbol{K}^{n}\boldsymbol{W}_{1}^{n},\label{eq:RK2_predictor}
\end{equation}
and the corrector evaluates the solution at time $(n+1)\D t$, 
\begin{eqnarray}
\boldsymbol{x}^{n+1} & = & \boldsymbol{x}^{n}+(1-w_{3}-w_{4})\Delta t\boldsymbol{L}^{n}\boldsymbol{x}^{n}+w_{3}\Delta t\boldsymbol{L}^{n}\tilde{\boldsymbol{x}}+w_{4}\Delta t\boldsymbol{L}^{n}\boldsymbol{x}^{n+1}\nonumber \\
 & + & w_{5}\Delta t(\widetilde{\boldsymbol{L}}-\boldsymbol{L}^{n})\tilde{\boldsymbol{x}}+w_{5}\Delta t\tilde{\boldsymbol{g}}+(1-w_{5})\Delta t\boldsymbol{g}^{n}\nonumber \\
 & + & \D t^{\frac{1}{2}}\left[(1-w_{6})\boldsymbol{I}+w_{6}\widetilde{\M M}\left(\M M^{n}\right)^{-1}\right]\M K^{n}\left[w_{2}^{\frac{1}{2}}\boldsymbol{W}_{1}^{n}+\left(1-w_{2}\right)^{\frac{1}{2}}\boldsymbol{W}_{2}^{n}\right].\label{eq:RK2_corrector}
\end{eqnarray}
The handling of the multiplicative noise term is inspired by the so-called
kinetic interpretation of the stochastic integral \cite{KineticStochasticIntegral_Ottinger}
and the well-known Fixman method for Brownian dynamics \cite{BD_Fixman}.
In the above discretization, the standard normal variates $\boldsymbol{W}_{1}^{n}$
correspond to the increment of the underlying Wiener processes over
the time interval $w_{2}\D t$, $\V{\mathcal{B}}\left(n\D t+w_{2}\D t\right)-\V{\mathcal{B}}\left(n\D t\right)=\left(w_{2}\Delta t\right)^{\frac{1}{2}}\boldsymbol{W}_{1}^{n}$,
while the normal variates $\boldsymbol{W}_{2}^{n}$ correspond to
the independent increment over the remainder of the time step, $\V{\mathcal{B}}\left(\left(n+1\right)\D t\right)-\V{\mathcal{B}}\left(n\D t+w_{2}\D t\right)=\left(\left(1-w_{2}\right)\Delta t\right)^{\frac{1}{2}}\boldsymbol{W}_{2}^{n}$.

\subsection{Additive Noise}

For additive noise, $\boldsymbol{K}(\boldsymbol{x})\equiv\boldsymbol{K}$,
we would like to achieve second-order weak accuracy. A second-order
integrator that uses derivatives is provided by the weak Taylor series
(\ref{eq:2nd_weak_derivatives}), in indicial notation with the implied
summation notation,
\begin{eqnarray}
x_{\alpha}\left(t+\D t\right) & = & x_{\alpha}\left(t\right)+\Delta tL_{\alpha\beta}^{n}x_{\beta}^{n}+\Delta tg_{\alpha}^{n}+K_{\alpha\beta}\Delta\mathcal{B}_{\beta}\nonumber \\
 & + & \left(\frac{\Delta t^{2}}{2}L_{\gamma\epsilon}^{n}x_{\epsilon}+\frac{\Delta t^{2}}{2}g_{\gamma}^{n}+\frac{\Delta t}{2}K_{\gamma\epsilon}\Delta\mathcal{B}_{\epsilon}\right)\left(x_{\beta}\partial_{\gamma}L_{\alpha\beta}^{n}+L_{\alpha\gamma}^{n}+\partial_{\gamma}g_{\alpha}^{n}\right)\nonumber \\
 & + & \frac{\Delta t^{2}}{4}K_{\gamma\epsilon}K_{\delta\epsilon}\left(x_{\beta}\partial_{\gamma}\partial_{\delta}L_{\alpha\beta}^{n}+2\partial_{\gamma}L_{\alpha\delta}^{n}+\partial_{\gamma}\partial_{\delta}g_{\alpha}^{n}\right)+O\left(\Delta t^{\frac{5}{2}}\right),\label{eq:RK1_derivatives}
\end{eqnarray}
where $\D{\V{\mathcal{B}}}=\V{\mathcal{B}}\left(\left(n+1\right)\D t\right)-\V{\mathcal{B}}\left(n\D t\right)$
is the Wiener increment, in law, $\D{\V{\mathcal{B}}}=\left(\Delta t\right)^{1/2}\boldsymbol{W}$
where $\boldsymbol{W}$ is a collection of i.i.d. standard normal
random variables.

To prove second-order order weak accuracy for the derivative-free
Runge-Kutta scheme, we need to match the first 5 moments of the numerical
increment $\Delta_{\alpha}^{n}=x_{\alpha}^{n+1}-x_{\alpha}^{n}$ to
the moments of the true increment $\Delta_{\alpha}(n\D t)=x_{\alpha}(\left(n+1\right)\D t)-x_{\alpha}(n\D t)$,
up to order $\Delta t^{2}$ \cite{MilsteinSDEBook}. The increment
of the RK method (\ref{eq:RK2_predictor},\ref{eq:RK2_corrector})
is (after recursively substituting in for $\boldsymbol{x}^{n+1}$,
and Taylor expanding certain terms) 
\begin{eqnarray*}
 & x_{\alpha}^{n+1} & =x_{\alpha}^{n}+\Delta tL_{\alpha\beta}^{n}x_{\beta}^{n}+\Delta tg_{\alpha}^{n}+\D t^{\frac{1}{2}}K_{\alpha\beta}\left[w_{2}^{\frac{1}{2}}\left(W_{1}^{n}\right)_{\beta}+\left(1-w_{2}\right)^{\frac{1}{2}}\left(W_{2}^{n}\right)_{\beta}\right]\\
 & + & \left(\Delta t^{2}L_{\gamma\epsilon}^{n}x_{\epsilon}^{n}+\Delta t^{2}g_{\gamma}^{n}\right)\left[w_{2}w_{5}x_{\beta}^{n}\partial_{\gamma}L_{\alpha\beta}^{n}+(w_{2}w_{3}+w_{4})L_{\alpha\gamma}^{n}+w_{2}w_{5}\partial_{\gamma}g_{\alpha}^{n}\right]\\
 & + & \Delta t^{3/2}K_{\gamma\epsilon}\left[w_{5}w_{2}^{\frac{1}{2}}x_{\beta}\left(\partial_{\gamma}L_{\alpha\beta}^{n}\right)\left(W_{1}^{n}\right)_{\epsilon}+\left((w_{3}+w_{4})w_{2}^{\frac{1}{2}}\left(W_{1}^{n}\right)_{\epsilon}+w_{4}\left(1-w_{2}\right)^{\frac{1}{2}}\left(W_{2}^{n}\right)_{\epsilon}\right)L_{\alpha\gamma}^{n}+w_{5}w_{2}^{\frac{1}{2}}\left(\partial_{\gamma}g_{\alpha}^{n}\right)\left(W_{1}^{n}\right)_{\epsilon}\right]\\
 & + & \frac{w_{2}w_{5}}{2}\Delta t^{2}K_{\gamma\epsilon}K_{\delta\epsilon}\left(x_{\beta}\partial_{\gamma}\partial_{\delta}L_{\alpha\beta}^{n}+2\partial_{\gamma}L_{\alpha\delta}^{n}+\partial_{\gamma}\partial_{\delta}g_{\alpha}^{n}\right)+O\left(\Delta t^{5/2}\right)
\end{eqnarray*}
Comparing the first moments of the increments, $E\left(\Delta_{\alpha}^{n}\right)-E\left(\Delta_{\alpha}(n\D t)\right)=O\left(\Delta t^{3}\right)$
if $w_{2}w_{5}=\frac{1}{2}$ and $w_{2}w_{3}+w_{4}=\frac{1}{2}$.
The difference in the second moments is also $O\left(\Delta t^{3}\right)$
under the same conditions, as is the difference in the third and fourth
moments. The fifth moments are already of $O\left(\Delta t^{3}\right)$.

\subsection{Multiplicative Noise}

For multiplicative noise, we only aim to achieve first-order weak
accuracy. For this, we need to match the first three moments of the
increment up to first order in $\D t$ \cite{MilsteinSDEBook}. With
a variable $\boldsymbol{K}(\boldsymbol{x})$, the difference in the
first moments obtains additional terms 
\[
(\alpha-w_{2}w_{6})\Delta t\,\partial_{\gamma}(K_{\alpha\beta}K_{\gamma\beta}^{n})+O(\Delta t^{2})
\]
Therefore, if we have $w_{2}w_{6}=\alpha$, then our method will be
consistent with the Ito SDE and reproduce the correct thermal drift
without explicitly evaluating $\partial_{\V x}\cdot\boldsymbol{M}(\boldsymbol{x})$.
Note however that this type of ``Fixman'' approach requires solving
one more linear system per time step because of the appearance of
$\left(\M M^{n}\right)^{-1}$ in (\ref{eq:RK2_corrector}).

Despite recent progress in mathematical understanding \cite{RoughBurgers_Hairer},
multiplicative noise has unclear physical relevance. Fluctuating hydrodynamics
is a coarse-grained model in which the variables are averages over
many microscopic degrees of freedom and therefore the fluctuations
are weak \cite{SmallNoiseWeak,CahnHilliard_WeakNoise,SDEMultistep_SmallNoise}.
To leading order in the magnitude of the fluctuations, the equations
are linear and the noise can be considered additive (though potentially
time-dependent). We speculate that multiplicative noise is a very
weak perturbation and its effects on the larger scales of the flow,
if any, can be sufficiently accurately captured by low-order temporal
integrators such as (\ref{eq:RK2_predictor},\ref{eq:RK2_corrector})
with the condition $w_{2}w_{6}=\alpha$.

\section{\label{sub:L0-stable}$L$-Stable Scheme}

In typical fluctuating hydrodynamics applications, for explicit schemes
the time step is severely limited not by advection but by momentum
or heat diffusion, notably, by viscous dissipation. For purely dissipative
linear equations, implicit handling of momentum diffusion can yield
$A$-stable schemes such as the implicit midpoint scheme (\ref{eq:Crank-Nicolson}).
This allows the use of much larger time step size $\D t$, at least
in principle. If one is interested in steady-state fluctuations, the
implicit midpoint scheme (\ref{eq:Crank-Nicolson}) gives the correct
spectrum of fluctuations for any $\D t$ (see the Appendix in Ref.
\cite{LLNS_Staggered} for a discussion of how to choose a suitable
$\D t$).

However, for time dependent linear problems, only an exponential integrator
can reproduce the correct dynamics for all modes (wavenumbers) for
all time step sizes. The implicit midpoint rule provides a notably
bad approximation to the exponential decay of correlations for large
$\D t$, since the Pade (1,1) rational approximation to the exponential
(\ref{eq:RatExp_mid}), $\exp(-x)\approx(1-x/2)/(1+x/2)$ tends to
-1 for $x\gg1$ instead of decaying to zero. This leads to oscillatory
dynamics for the modes that are under-resolved by the large time step
size, i.e., for the thermal fluctuations at large wavenumbers. A much
better approximation to $\exp(-x)$ is provided by rational approximations
that decay to zero as $x\rightarrow\infty$. In numerical analysis
jargon this means handling the diffusive fluxes using an $L$-stable
numerical method.

\begin{figure}
\begin{centering}
\includegraphics[width=0.5\textwidth]{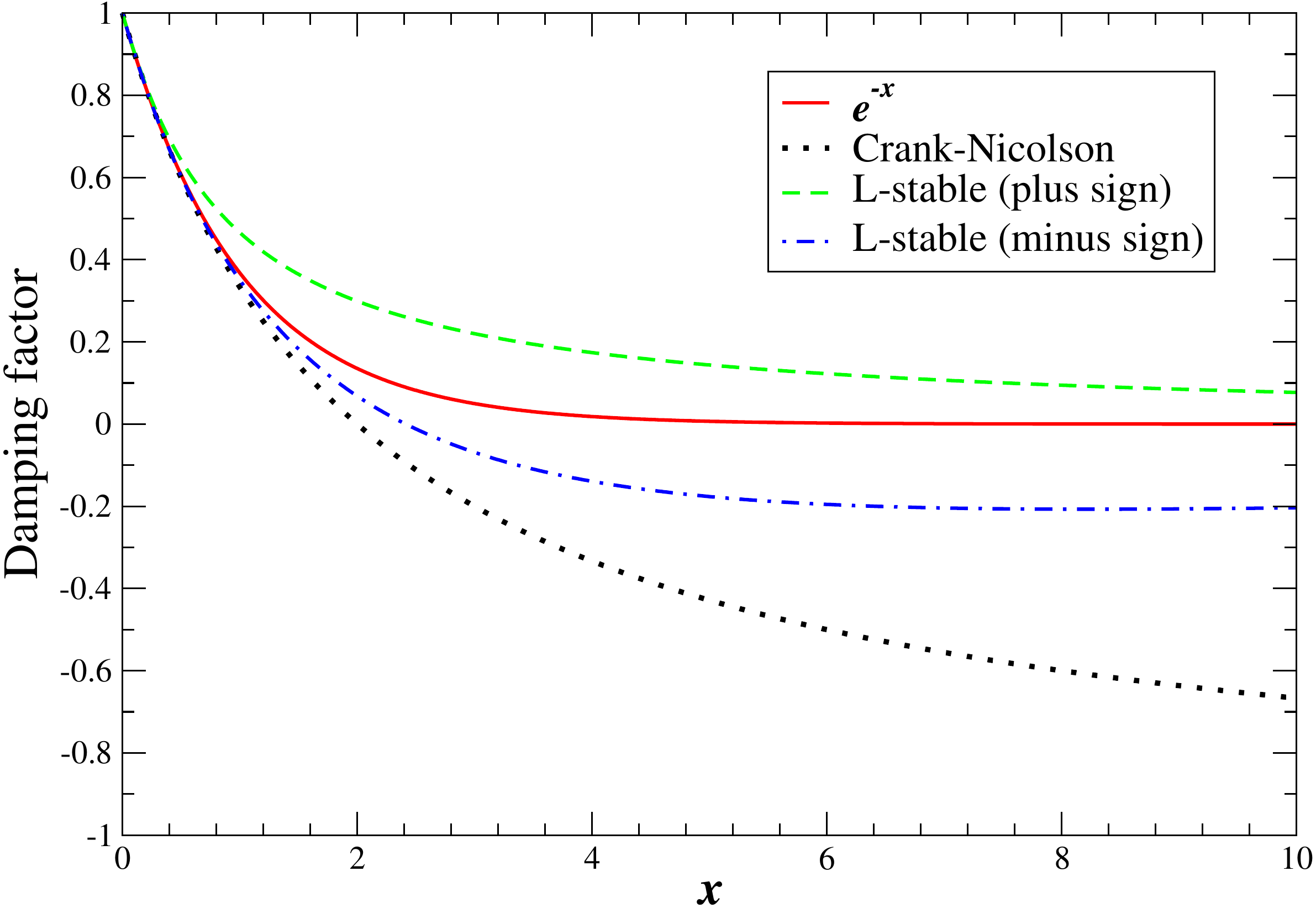}
\par\end{centering}

\caption{\label{fig:RationalApprox_exp}Comparison of $\exp(-x)$ with three
rational approximations. The approximation in the Crank-Nicolson scheme
(\ref{eq:RatExp_mid}) does not decay to zero for large $x$. The
approximation (\ref{eq:RatExp_L0}) decays to zero for $w_{4}=1\pm\sqrt{2}/2$,
however, only the positive sign gives a strictly positive approximation.}
\end{figure}

Let us consider the choice of weights in the general scheme (\ref{eq:generic_RK2})
that yield a scheme that is weakly second-order accurate and $L$-stable
in the implicit part of the dynamics. From the conditions for second-order
accuracy (\ref{eq:conditions_2nd}) we obtain 
\[
w_{3}=\frac{\frac{1}{2}-w_{4}}{w_{2}},\quad w_{5}=\frac{1}{2w_{2}},
\]
and from the condition of $L$ stability we obtain 
\[
w_{1}=\frac{\frac{1}{2}-w_{4}}{1-w_{4}},
\]
which gives the following rational approximation to the exponential
decay of the dynamics,
\begin{equation}
\exp\left(-x\right)\approx\frac{\left(1-2\, w_{4}+2\, w_{4}^{2}\right)x-2\left(1-w_{4}\right)}{\left(w_{4}\, x+1\right)\left[\left(2\, w_{4}-1\right)x-2\left(1-w_{4}\right)\right]}.\label{eq:RatExp_L0}
\end{equation}
A reasonable choice of $w_{4}$ can be taken to be the one that minimizes
the mismatch between the coefficient in front of $x^{3}$ in the Taylor
series expansion of the left and right hand sides \cite{L_stable_HeatEq},
giving $w_{4}=1\pm\sqrt{2}/2$. In Fig. \ref{fig:RationalApprox_exp}
we compare the two rational approximations by choosing the plus or
minus sign. In the deterministic literature the choice of the minus
sign has been favored \cite{L_stable_HeatEq,IMEX_PDEs}, however,
we recommend the plus sign, 
\[
w_{4}=1+\frac{\sqrt{2}}{2},
\]
because this gives a strictly positive approximation to the exponential
decay of correlations instead of oscillatory behavior for the under-resolved
modes (large $x$).

We are still left with the choice of $w_{2}$, where the two common
choices for $w_{2}$ would be a mid-point, $w_{2}=1/2$, or an end-point,
$w_{2}=1$, predictor stage. From the discussion in Section \ref{sub:Explicit-Midpoint}
we know that when all terms are handled explicitly ($\M L=\M 0$),
the choice $w_{2}=1/2$ gives third-order accuracy for the static
covariance for linear problems. It can also be shown that this choice
leads to third-order accuracy of the static covariances in the linearized
setting if all terms are discretized implicitly ($\boldsymbol{g}(\boldsymbol{x})=\V 0$).
This suggests that a better, even if not unique, choice, is to take
$w_{2}=1/2$, giving our preferred choice of weights for an $L$-stable
predictor-corrector scheme,
\begin{equation}
w_{1}=w_{4}=1+\frac{\sqrt{2}}{2},\, w_{2}=\frac{1}{2},\, w_{3}=-(1+\sqrt{2}),\, w_{5}=1.\label{eq:L0_stable}
\end{equation}
In the linearized setting, this $L$-stable scheme gives second-order
accurate covariances for small time step sizes; however, it does not
produce the correct spectrum for the fluctuations for \emph{large}
time step sizes, unlike the implicit midpoint scheme (\ref{eq:Crank-Nicolson}).
In particular, for large $\D t$ the $L$-stable scheme strongly damps
the magnitude of the fluctuations of the fast (small wavelength or
large wavenumber) modes. Therefore, if static covariances are the
quantity of interest, the implicit midpoint rule should be used instead.

\section{\label{sec:ApproxProj}Approximate Projection Methods}

Here we consider a generalization of the projected Euler-Maruyama
scheme (\ref{eq:projected_EM_exact}), 
\begin{equation}
\V v^{n+1}=\widetilde{\M{\Set P}}\left[\V v^{n}+\nu\D t\,\M L_{\V v}\V v^{n}+\left(2\nu\D t\right)^{\frac{1}{2}}\M D_{\V w}\V W_{\V v}^{n}\right],\label{eq:projected_EM_approx}
\end{equation}
where $\widetilde{\M{\Set P}}$ an approximation to the discrete projection
$\M{\Set P}$, for example, $\M{\Set P}=\M I-\M G\M L_{p}^{-1}\M D$.
Here $\M L_{p}$ is a discrete pressure Laplacian operator that may,
in general, be different from $\M L_{s}=\M D\M G$. For example, with
spatial discretizations of the incompressible (Navier-)Stokes equations
that use cell-centered velocities, $\M L_{s}$ possesses a non-trivial
nullspace and the corresponding exact projection methods (in which
$\widetilde{\M{\Set P}}=\M{\Set P}$) suffer from the so-called checkerboard
instability. Approximate projection methods have been developed to
overcome these difficulties of exact cell-centered projection methods
\cite{almgrenBellSzymczak:1996}. One of the simplest approximate
projection methods is (\ref{eq:projected_EM_approx}) with $\M L_{p}$
being the standard second-order Laplacian stencil \cite{ApproximateProjection_I}.
For the staggered-grid spatial discretization we employ, however,
it is straightforward to invert $\M L_{s}$ and approximate projection
methods are not used in practice.

The steady-state covariance of the iteration (\ref{eq:projected_EM_approx})
should be a consistent approximation to the continuum result (\ref{eq:C_v_cont}).
Specifically, we ask that to leading order in the time step size
\[
\M C_{\V v}=\av{\V v^{n+1}\left(\V v^{n+1}\right)^{\star}}=\av{\V v^{n}\left(\V v^{n}\right)^{\star}}=\M{\Set P}+\D t\,\D{\M C}_{\V v}+O\left(\D t^{2}\right).
\]
Substituting (\ref{eq:projected_EM_approx}) in this condition and
equating the leading-order terms we obtain the condition
\[
\widetilde{\M{\Set P}}\M{\Set P}\widetilde{\M{\Set P}}^{\star}=\M{\Set P}.
\]
This condition is satisfied for \emph{exact projection} methods, $\widetilde{\M{\Set P}}=\M{\Set P}$,
but not for approximate projection methods, $\widetilde{\M{\Set P}}\neq\M{\Set P}$.
Assuming the initial condition is discretely divergence-free, for
exact projection (\ref{eq:projected_EM_approx}) is equivalent to
(\ref{eq:projected_EM_exact}).

\section{\label{AppendixModeAnalysis}Mode Analysis}

It is instructive to describe a framework for analyzing schemes such
as (\ref{eq:incompressible_CN_linear}), following the mode analysis
used to study splitting errors in projection methods in the deterministic
context \cite{ProjectionModes_III,GaugeIncompressible_E}. This analysis
can in principle produce explicit expressions for the spectrum of
velocity fluctuations for the types of schemes we consider here. It
also illustrates clearly the role of the pressure and, in particular,
the difficulties with applying semi-\emph{implicit} projection (splitting)
methods in the context of the fluctuating Navier-Stokes equations.

A mode of the spatially-discretized unforced time-dependent (creeping)
Stokes flow equation
\begin{equation}
\partial_{t}\V v+\M G\V{\pi}=\nu\M L_{\V v}\V v,\text{ s.t. }\M D\V v=0\label{eq:spatial_Stokes}
\end{equation}
is an exponentially-decaying solution of the form
\[
\V v\left(t\right)=\V v_{0}e^{-\sigma t}\mbox{ and }\V{\pi}\left(t\right)=\nu\V{\pi}_{0}e^{-\sigma t}.
\]
Here $\sigma\geq0$ is the decay rate associated with the spatial
mode $\left\{ \V v_{0},\,\V{\pi}_{0}\right\} $, which is a normalized
solution to the eigen-problem
\begin{equation}
\left(\M L_{\V v}+\nu^{-1}\sigma\M I\right)\V v_{0}+\M G\V{\pi}_{0}=0\text{ and }\M D\V v_{0}=0.\label{eq:continuum_modes}
\end{equation}
These modes diagonalize the creeping Stokes flow dynamics and form
a complete orthonormal basis for the space of divergence-free velocity
fields. This can be seen by eliminating the pressure to obtain the
classical eigenvalue problem in the subspace of discretely divergence-free
velocity fields
\[
\left[\M L_{\V v}-\M L_{\V v}^{-1}\M G\left(\M D\M L_{\V v}^{-1}\M G\right)^{-1}\M D\M L_{\V v}^{-1}\right]\V v_{0}=-\nu^{-1}\sigma\V v_{0}.
\]

In the presence of a stochastic forcing, we can express any solution
in a basis formed by the modes $\left\{ \V v_{0}^{1},\V v_{0}^{2},\dots\right\} $,
\[
\V v\left(t\right)=\sum_{k}\upsilon_{k}(t)\,\V v_{0}^{k},
\]
where the mode amplitudes $\upsilon_{k}(t)$ are scalar stochastic
processes. The stochastic forcing $\left(2\nu\right)^{\frac{1}{2}}\M D_{\V w}\V{\mathcal{W}}_{\V v}\left(t\right)$
in the momentum equation can be projected onto $\V v_{0}^{k}$ to
obtain the amplitude of the stochastic forcing for mode $k$,
\[
w_{k}\left(t\right)=\left(2\nu\right)^{\frac{1}{2}}\left(\V v_{0}^{k}\right)^{\star}\left(\M D_{\V w}\V{\mathcal{W}}_{\V v}\left(t\right)\right),
\]
which is a scalar white-noise process with covariance
\[
\av{w_{k}\left(t\right)w_{k}^{\star}\left(t^{\prime}\right)}=2\nu\left(\V v_{0}^{k}\right)^{\star}\left[\M D_{\V w}\av{\V{\mathcal{W}}_{\V v}\left(t\right)\V{\mathcal{W}}_{\V v}^{\star}\left(t^{\prime}\right)}\M D_{\V w}^{\star}\right]\left(\V v_{0}^{k}\right)=-2\nu\left(\V v_{0}^{k}\right)^{\star}\M L_{\V v}\left(\V v_{0}^{k}\right)\,\delta\left(t-t^{\prime}\right),
\]
where we made use of the discrete fluctuation-dissipation balance
between the viscous dissipation and the stochastic forcing, $\M L_{\V v}=-\M D_{\V w}\left(\M D_{\V w}\right)^{\star}$.
From (\ref{eq:continuum_modes}) we can express
\[
-2\nu\left(\V v_{0}^{k}\right)^{\star}\M L_{\V v}\left(\V v_{0}^{k}\right)=-2\nu\left(\V v_{0}^{k}\right)^{\star}\left(\M G\V{\pi}_{0}-\nu^{-1}\sigma_{k}\V v_{0}^{k}\right)=2\nu\left(\M D\V v_{0}^{k}\right)^{\star}\V{\pi}_{0}+2\sigma_{k}\norm{\V v_{0}^{k}}^{2}=2\sigma_{k},
\]
where we again made use of the duality relation $\M G=-\M D^{\star}$.
This simple calculation shows that in the mode representation the
linearized fluctuating Navier-Stokes equation becomes a collection
of decoupled scalar Langevin equations driven by standard Wiener processes,
\begin{equation}
\frac{d\upsilon_{k}}{dt}=-\sigma_{k}\upsilon_{k}+\left(2\sigma_{k}\right)^{\frac{1}{2}}\mathcal{W}_{\upsilon}(t).\label{eq:modes_SDEs}
\end{equation}
The fluctuation-dissipation balance is most clearly revealed in this
representation.

Temporal discretizations can be analyzed by projecting the numerical
solution onto a set of \emph{discrete} modes. For the implicit midpoint
discretization (\ref{eq:incompressible_CN_linear}), the modes are
decaying solutions of the form 
\[
\V v_{k}^{n+1}=\V v_{k}^{n}e^{-\tilde{\sigma}_{k}\D t}\text{ and }\V{\pi}_{k}^{n+\frac{1}{2}}=\nu\V{\pi}_{k}^{n}e^{-\tilde{\sigma}_{k}\D t/2},
\]
where $\tilde{\sigma}_{k}\approx\sigma_{k}$ is the numerical decay
rate. The spatial structure of the mode $\left\{ \V v_{k}^{n},\V{\pi}_{k}^{n}\right\} $
is the solution to the discrete eigen-problem
\[
\left[\M L_{\V v}+\nu^{-1}\left(\frac{2}{e^{-\tilde{\sigma}_{k}\D t}+1}\right)\left(\frac{1-e^{-\tilde{\sigma}_{k}\D t}}{\D t}\right)\M I\right]\V v_{k}^{n}+\left(\frac{2}{e^{-\tilde{\sigma}_{k}\D t}+1}\right)e^{-\tilde{\sigma}_{k}\D t/2}\,\M G\V{\pi}_{k}^{n}=\V v^{n}.
\]
Comparison to (\ref{eq:continuum_modes}) shows that the spatial modes
are the same as for the semi-continuum (\ref{eq:spatial_Stokes}),
and the temporal decay rate is second-order accurate in the time step,
\[
\sigma_{k}=\left(\frac{2}{e^{-\tilde{\sigma}_{k}\D t}+1}\right)\left(\frac{1-e^{-\tilde{\sigma}_{k}\D t}}{\D t}\right)=\tilde{\sigma}_{k}\left(1-\frac{\tilde{\sigma}_{k}^{2}\D t^{2}}{12}\right)+O\left(\D t^{3}\right).
\]
When the stochastic forcing is included, the discrete velocity can
be represented in the basis formed by the discrete modes just as we
did above for the time-continuous equations. In the mode representation
the scheme (\ref{eq:incompressible_CN_linear}) is seen to be nothing
more than the implicit midpoint method (\ref{eq:Crank-Nicolson})
applied to the system of decoupled SDEs (\ref{eq:modes_SDEs}).

The mode analysis reveals that semi-implicit projection (splitting)
methods have a significant shortcoming not seen for explicit methods.
A Crank-Nicolson projection method for (\ref{eq:spatial_Stokes})
consists of first solving the following linear system for the velocity
$\M v^{n+1}$ with a time-lagged pressure \cite{bellColellaGlaz:1989},
\[
\left(\M I-\frac{\nu\D t}{2}\right)\tilde{\V v}^{n+1}+\D t\,\M G\V{\pi}^{n-\frac{1}{2}}=\left(\M I+\frac{\nu\D t}{2}\right)\V v^{n},
\]
and then projecting the intermediate velocity $\tilde{\V v}^{n+1}$
to enforce the divergence-free constraint, $\V v^{n+1}=\M{\Set P}\tilde{\V v}^{n+1},$
by solving a linear system for the pressure correction
\[
\V v^{n+1}=\tilde{\V v}^{n+1}-\D t\,\M G\D{\V{\pi}}^{n},\text{ s.t. }\M D\V v^{n+1}=0.
\]
Repeating the discrete mode calculation reveals that the spatial modes
for the above temporal discretization are not the same as for the
semi-continuum (\ref{eq:spatial_Stokes}), specifically, the gradient
of pressure term in (\ref{eq:continuum_modes}) is modified by a term
involving the Laplacian $\M L_{\V v}$. For periodic systems the discrete
gradient and vector Laplacian commute, $\M L_{\V v}\M G=\M G\M L_{s}$,
and modes have the correct spatial structure. However, for non-periodic
systems the splitting of the pressure and velocity equations introduces
a commutator error that leads to the appearance of ``spurious''
or ``parasitic'' modes \cite{GaugeIncompressible_E}. For deterministic
solutions and moderate time step sizes, spatio-temporal smoothness
of the solution usually makes these commutator errors acceptably small.
In the stochastic context, however, all modes are stochastically forced
and have a non-negligible amplitude, including the parasitic modes.
For this reason, we chose to use (\ref{eq:incompressible_CN_linear})
and solve a coupled Stokes linear system for \emph{both} pressure
and velocity, and only use the projection method as a preconditioner
for the required Krylov solver \cite{NonProjection_Griffith}. We
emphasize again that for purely explicit time stepping scheme the
spatial structure of the modes is preserved and projection methods
can be used in the stochastic setting as well.

\end{appendix}


\end{document}